\documentclass[10pt,journal,a4paper,twoside,twocolumn]{IEEEtran}
\usepackage{cite}
\usepackage{float}
\usepackage{stfloats}
\usepackage{fancyhdr}
\usepackage{color}
\usepackage[inline]{enumitem}
\usepackage{amsmath,amsfonts,amssymb}
\usepackage{graphicx}
\usepackage{algorithm}
\usepackage{algorithmic}
\usepackage{epsfig}
\usepackage{hyperref}
\usepackage{epstopdf}
\usepackage{bm}
\usepackage{array}

\usepackage{bbm}
\usepackage{pifont}
\usepackage{tikz}
\usepackage{multirow}
\usepackage{bbding}
\usepackage[caption=false]{subfig}
\usetikzlibrary{positioning}

\newtheorem{Lemma}{Lemma}
\newtheorem{Rem}{Remark}

\begin{document}

	\title{Algorithm Design and Prototype Validation for Reconfigurable Intelligent Sensing Surface: Forward-Only Transmission
	}
	\author{\IEEEauthorblockN{
		Cheng Luo, \IEEEmembership{Graduate Student Member, IEEE}, Luping Xiang, \IEEEmembership{Senior Member, IEEE}, Jie Hu, \IEEEmembership{Senior Member, IEEE} and Kun Yang, \IEEEmembership{Fellow, IEEE}
		}
			\\

		\thanks{Cheng Luo and Jie Hu are with the School of Information and Communication Engineering, University of Electronic Science and Technology of China, Chengdu, 611731, China, email: chengluo@std.uestc.edu.cn; hujie@uestc.edu.cn.}
		\thanks{Luping Xiang and Kun Yang are with the State Key Laboratory of Novel Software Technology, Nanjing University, Nanjing 210008, China, and School of Intelligent Software and Engineering, Nanjing University (Suzhou Campus), Suzhou, email: luping.xiang@uestc.edu.cn; kunyang@essex.ac.uk.}
		\thanks{This paper was funded in part by the Natural Science Foundation of China under Grant 62301122, in part by the Natural Science Foundation of China under Grant 62132004, in part by the Quzhou Government under Grant 2024D007 and 2023D005, in part by the Nanjing University - China Mobile Communications Group Co., Ltd. Joint Institute, in part by the Gusu Innovation Project (Grant No.: ZXL2024360), in part by the Jiangsu Major Project on Fundamental Research (Grant No.: BK20243059) and in part by the High-Tech District of Suzhou City (Grant No.: RC2025001) (Corresponding author: Luping Xiang.)}	 
	}

	\maketitle

	\thispagestyle{fancy} 
	\lhead{} 
	\chead{} 
	\rhead{} 
	\lfoot{} 
	\cfoot{} 
	\rfoot{\thepage} 
	\renewcommand{\headrulewidth}{0pt} 
	\renewcommand{\footrulewidth}{0pt} 
	\pagestyle{fancy}

    \rfoot{\thepage} 

	\begin{abstract}
	Sensing-assisted communication schemes have recently garnered significant research attention. In this work, we design a dual-function reconfigurable intelligent surface (RIS), integrating both active and passive elements, referred to as the reconfigurable intelligent sensing surface (RISS), to enhance communication. By leveraging sensing results from the active elements, we propose communication enhancement and robust interference suppression schemes for both near-field and far-field models, implemented through the passive elements. These schemes remove the need for base station (BS) feedback for RISS control, simplifying the communication process by replacing traditional channel state information (CSI) feedback with real-time sensing from the active elements. The proposed schemes are theoretically analyzed and then validated using software-defined radio (SDR). Experimental results demonstrate the effectiveness of the sensing algorithms in real-world scenarios, such as direction of arrival (DOA) estimation and radio frequency (RF) identification recognition. Moreover, the RISS-assisted communication system shows strong performance in communication enhancement and interference suppression, particularly in near-field models.
	\end{abstract}
\section{Introduction}

\IEEEPARstart{R}{econfigurable} intelligent surfaces (RIS) represent a promising innovation for future sixth-generation (6G) networks. This technology consists of a planar surface embedded with numerous cost-effective, passive reflecting elements, each capable of independently adjusting the phase and amplitude of incident electromagnetic signals. This unique capability allows for customized optimization to meet specific functional and performance demands, effectively mitigating the challenges posed by complex wireless propagation environments. Consequently, RIS enhances communication and energy transfer efficiency, coverage, and security\cite{wqqtutorial, LYWtutorial}. Moreover, the introduction of an additional line-of-sight (LoS) channel further improves signal reliability and transmission efficiency.

Due to its low power consumption and simplified hardware architecture, RIS has gained widespread studies in integrated sensing and communication (ISAC) systems\cite{RIS_ISAC,RIS_ISAC2, JointCommSensing} and simultaneous wireless information and power transfer (SWIPT) systems \cite{RISSWIPT3, RISSWIPT4}. Moreover, their utility extends to various innovative domains, including robust beamforming techniques \cite{robustBF1}, electromagnetic wave frequency mixing \cite{frequencyadjustment2}, active RIS configurations \cite{activeIRS2}, and localization \cite{RISS_loc}, providing significant advantages in these areas.

However, several challenges remain, particularly in acquiring accurate channel state information (CSI) between the RIS and its associated primary base stations (BS) or users\cite{CSIchallenge3, CSIchallenge, CSIchallenge2}. The passive nature of RIS components complicates the CSI acquisition process, resulting in increased pilot overhead that scales with the number of RIS elements and users, as highlighted in\cite{pilotoverhead1, pilotoverhead2}, thereby complicating conventional channel estimation methods.

Recent studies have sought to address these challenges by developing innovative methods for RIS channel estimation. \cite{CS_CE} proposes a compressed sensing approach for estimating RIS cascaded channels, significantly reducing the channel overhead. Another novel solution involves placing power sensors behind each RIS element to enable phase-coherent signal superposition at the receiver through interference observation\cite{sensingirs_dll}. A location-based information systems have been employed to enhance beamforming strategies, and detailed performance analysis are provided in \cite{locinformation}. Additionally, empirical data has been leveraged to optimize real world RIS precoding and phase adjustments without knowledge of CSI\cite{blindBF}. A recent study further outlines a CSI-free approach for RIS-assisted large-scale wireless energy transfer (WET), efficiently directing energy across spatial domains using a beam rotation strategy\cite{Luo_MassiveWE}.

Moreover, numerous studies have focused on sensing-assisted communication, exploring the potential for collaboration on channel estimation through sensing results or using sensing as an alternative to traditional channel estimation methods. This approach presents both new opportunities and challenges. Specifically, limited CSI is utilized for communication and rough sensing during the initial phase. The collected sensing information is then used to enhance both communication and sensing in the subsequent phase\cite{twophase}. This two-phase ISAC transmission protocol for RIS-aided MIMO ISAC systems enables cooperative operation between the two functions. Similarly, \cite{SensingAidedComm3} presents two deep learning networks designed for beam selection and power allocation, respectively. With the assistance of sensing data, these networks work together to maximize the communication channel capacity. 

Furthermore, a RIS structure combining active and passive elements, referred to as reconfigurable intelligent sensing surface (RISS), was proposed to enhance sensing capabilities in \cite{9724202}. Building on this architecture, a joint time-of-arrival (TOA) and direction-of-arrival (DOA) estimation method for localization was developed, along with the derivation of the corresponding Cramér–Rao bound (CRB) \cite{RISS_loc}, further demonstrating the potential of RISS in sensing tasks. In addition, \cite{RISS_luo} proposed a RISS-assisted wireless powered communication network (WPCN), where the active elements are employed to sensing the DOA of the BS and the users in the uplink and downlink, respectively. The sensing results are then utilized to facilitate uplink wireless information transfer (WIT) and downlink WET. Similarly, \cite{10497119} introduced a multi-RISS-assisted ISAC framework, which jointly optimizes transmit and reflective beamforming to minimize the worst-case CRB for target estimation while guaranteeing communication quality. More recently, \cite{chen2025integrated} developed a space-time-coding metasurface that is capable of real-time incident DOA sensing. The proposed coding matrix dynamically adapts the beam direction, enabling coordinated and real-time operation of communication and sensing functions. Partial related works are presented in Table \ref{table:relatedworks11}.
\begin{table*}[]
    \centering
	\caption{Relevant works on RIS-assisted communicaton with sensing.}
	\begin{tabular}{c|c|c|c|c|c}
    \hline
    \textbf{} &\textbf{Sensing scheme}& \textbf{Comm.} &\textbf{Interf. suppr.} &\textbf{Robust design} & \textbf{Implementation} \\
    \hline
    Our proposed&DOA estimation, signal detection&\CheckmarkBold&\CheckmarkBold&\CheckmarkBold&\CheckmarkBold\\
    \hline
	GPS-based\cite{locinformation}&GPS location&\CheckmarkBold& &\CheckmarkBold& \\
	\hline
	Two-phase scheme\cite{twophase}&Two-phase sensing &\CheckmarkBold& & & \\
	\hline
	Two-network scheme\cite{SensingAidedComm3}&Radar sensing&\CheckmarkBold& & & \\
	\hline
	RISS-sensing\cite{9724202}&DOA estimation& & & &\\
	\hline
	RISS-SWIPT\cite{RISS_luo}&DOA estimation&\CheckmarkBold& &\CheckmarkBold& \\
	\hline
	RISS-location\cite{RISS_loc}&DOA estimation& & & & \\
	\hline
	RISS-ISAC\cite{10497119}&DOA estimation&\CheckmarkBold& & & \\
	\hline
	STCM\cite{chen2025integrated}&DOA estimation&\CheckmarkBold& & &\CheckmarkBold\\
	\hline
    \end{tabular} \label{table:relatedworks11}
\end{table*}

As a step ahead from \cite{RISS_luo}, we propose a communication enhancement scheme by sensing, including target enhancement and robust interference suppression. And then we design the hardware of RISS for evaluation. Our contributions are summarized as follows:

\begin{itemize}
	\item We extend the conventional sensing paradigm of RISS by proposing a dual-sensing enhanced communication architecture. In this framework, active sensing elements within the RISS are utilized for both DOA estimation and RF identification, enabling effective discrimination between target and interference signals. 
	\item Based on the acquired sensing information, we design the reflection phase shift matrix for the passive elements. Both near-field and far-field communication channel models are considered. Furthermore, robust beamforming strategies are developed to enhance the target signal and suppress interference, taking into account potential sensing errors, thereby significantly improving the signal-to-interference-plus-noise ratio (SINR) at the receiver.
	\item By leveraging the real-time sensing capability of RISS, the proposed system eliminates the reliance on base station (BS) control and reduces the complexity of traditional CSI feedback. Consequently, the RISS can be directly integrated into existing communication systems and operate in a purely forward transmission manner, which simplifies the overall communication process.
	\item To validate the proposed framework, a prototype hardware system is developed, consisting of a 2-bit RIS with 100 passive elements and a uniform linear array (ULA) with 4 active elements. We implement two key sensing algorithms—DOA estimation and RF identification—along with the communication module on a software-defined radio (SDR) platform. Extensive experiments are conducted to evaluate the performance of the proposed target enhancement and interference suppression schemes. The results confirm the effectiveness of our approach and lay a solid foundation for future research on ISAC beamforming and system design.
\end{itemize}

The remainder of this paper is organized as follows: Section \ref{sec:II} presents an overview of the system model, covering both the near-field and far-field models. Section \ref{sec:III} discusses the fundamentals of sensing-assisted communication. The hardware and experimental details are provided in Section \ref{sec:IV}. Experimental results are outlined in Section \ref{sec:V}, followed by the findings and conclusions in Section \ref{sec:VI}.

\emph{Notation:} $\mathbf{I}_{M}$ denotes the $M \times M$ identity matrix, and $\mathbf{1}_{M}$ represents the $M \times 1$ column vector of all ones. The notation $[\cdot]_i$ refers to the $i$-th element of a vector, while $[\cdot]_{i,j}$ denotes the $(i,j)$-th element of a matrix. The imaginary unit is represented by $\mathbbm{i} = \sqrt{-1}$. The Euclidean norm is denoted by $||\cdot||$, and the absolute value is denoted by $|\cdot|$. The function $\text{diag}(\cdot)$ constructs a diagonal matrix. The operators $(\cdot)^{T}$ and $(\cdot)^{H}$ denote the transpose and conjugate transpose, respectively. Finally, $\mathcal{CN}$ indicates the circularly symmetric complex Gaussian distribution.

\begin{figure}
	\centering
	\includegraphics[width=0.90\linewidth]{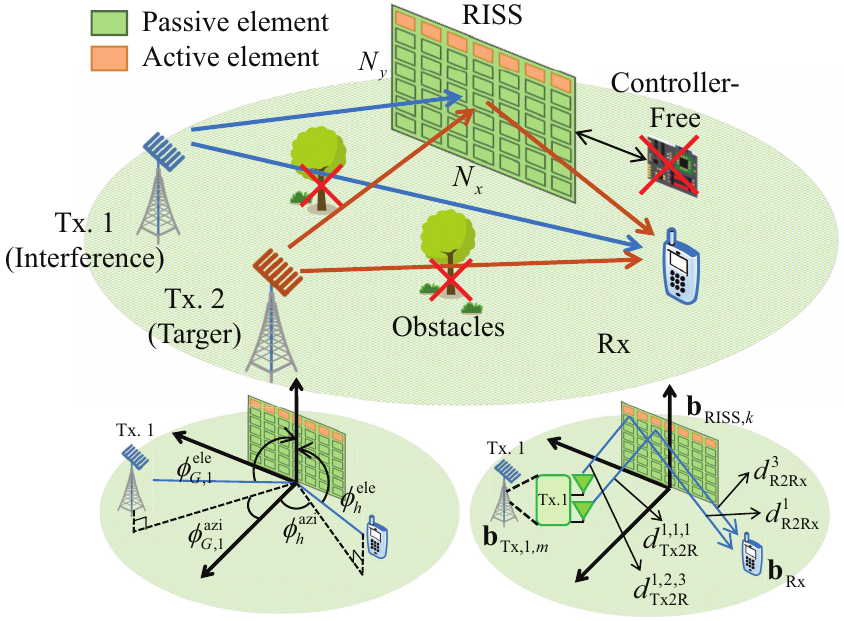}
	\caption{System model of proposed RISS-assisted communication scheme. The RISS is composed of both active and passive elements.}
	\label{fig:systemmodel}
\end{figure}
\section{System Model}\label{sec:II}
Far-field and near-field channel models are both adopted for our system in this paper. We assume two $M$ antennas transmitters (Tx) transmit signal simultaneously to a single antenna receiver (Rx) via the RISS, which consists of total $(N_a+N)$ elements, including $N_a$ active elements and $N=N_x\times N_y$ passive elements, as depicts in Fig. \ref{fig:systemmodel}. We assume these active elements involves the RISS have the ability of signal processing, e.g., DOA estimation and identification recognition. This enables the RISS to operate independently of the BS, eliminating the need for feedback control from the Rx (as shown in Fig. \ref{fig:systemmodel}), and allows the passive phase shift matrix of RISS to be designed based on sensing information, without relying on traditional CSI feedback. Additionally, there is no direct link between the Rx and Tx.1 and Tx.2 due to blockages, without loss of generality\footnote{In fact, the orthogonal polarization between the incident and reflected signals of the RIS effectively blocks the direct link between the Tx and Rx in Section \ref{sec:IV}.}.

\begin{Rem}
	The placement of both active and passive elements in the RISS plays a crucial role in the proposed framework. In this work, the active elements are arranged in a ULA (or an L-array for two-dimensional DOA estimation), along one side of the RISS, as illustrated in Fig. \ref{fig:systemmodel}. This configuration is adopted for the following reasons:
    \begin{enumerate*} 
        \item DOA estimation algorithms based on ULA and L-arrays have been extensively studied and validated.
        \item The passive elements retain a complete uniform planar array (UPA) structure, which allows for the direct application of conventional RIS beamforming and optimization methods. 
        \item The spatial separation between active and passive elements helps mitigate potential hardware coupling effects.
    \end{enumerate*}

	Moreover, we advocate for a moderate increase in the number of active elements in the RISS, rather than an unbounded expansion. This recommendation is supported by two key observations. First, the reflected beam inherently has a finite beamwidth, which provides robustness to DOA estimation errors and reduces the system’s sensitivity to sensing inaccuracies. Second, a larger number of passive elements provides greater degrees of freedom for interference suppression, thereby enhancing system robustness against angular deviations in sensing.
	
	Additionally, increasing the number of active elements incurs significant hardware costs, including more RF chains and higher power consumption, making large-scale deployment impractical in many scenarios. Therefore, a balanced design should aim to ensure sufficient sensing capability with a limited number of active elements, while maximizing communication performance through an expanded passive array.
	
	The relationship between sensing accuracy and communication performance has been analyzed in \cite{RISS_luo}, offering further insights into the influence of sensing precision on system performance.
\end{Rem}

\subsection{Far-field Model}
Quasi-static block line-of-sight (LoS) channels are taken into account\footnote{It is worth noting that the proposed scheme for target signal enhancement and interference suppression based on the LoS channel can be directly extended to more complex Rician channels. As shown in \cite{RISS_luo}, in scenarios where the LoS component is predominant, relying solely on LoS-based sensing enables near-optimal communication performance relative to full CSI, with substantially lower pilot overhead.}. Specifically, the channel between the Tx-to-RISS and RISS-to-Rx are denoted as $\mathbf{G}_i\in\mathbb{C}^{N\times M},\forall i=\{1,2\}$ and $\mathbf{h}\in\mathbb{C}^{N\times 1}$, then we have
\begin{align}
	&\mathbf{G}_i=\sqrt{MN} \boldsymbol{\alpha}\left(\vartheta_{G,i}, \varphi_{G,i}\right) \boldsymbol{\beta}^{\mathrm{T}}\left(\varpi_{G,i}\right),\forall i\in \{1,2\}, \label{eqn:farfG}\\
	&\mathbf{h}=\sqrt{N} \boldsymbol{\alpha}\left(\vartheta_h, \varphi_h\right), \label{eqn:farfh}\\
	&\boldsymbol{\beta}(\varpi_{G,i})=\frac{1}{\sqrt{M}}\left[1, e^{\mathrm{i} \varpi_{G,i}},e^{\mathrm{i}2 \varpi_{G,i}}, \cdots, e^{\mathrm{i} (M-1)\varpi_{G,i}}\right], \\
	&\boldsymbol{\alpha}(\vartheta_X, \varphi_X)=\boldsymbol{\alpha}_x\left(\vartheta_X\right)\otimes \boldsymbol{\alpha}_y\left(\varphi_X\right), X\in\left\{h, (G,i)\right\}
\end{align}
where $\vartheta_X=2\pi d/\lambda\cos(\phi_X^{\text{ele}})=\pi\cos(\phi_X^{\text{ele}})$ by setting $d/\lambda=1/2$ without loss of generality\footnote{Excessively large $d$ can introduce grating lobes, leading to energy dispersion and distortion of the radiation pattern, while excessively small $d$ may cause mutual coupling between antennas. Considering these factors, setting $d=\lambda/2$ is a widely accepted compromise that balances both effects.}. $d$ and $\lambda$ represent the element spacing and carrier wavelength, respectively. Similarly, we express $\varphi_X=2\pi d/\lambda \sin(\phi^\text{ele}_X)\cos(\phi^\text{azi}_X)=\pi\sin(\phi^\text{ele}_X)\cos(\phi^\text{azi}_X)$ and $\varpi_{G,i} = \pi\sin(\phi_{G,i}^{\text{dep}})$\cite{9103231, 9724202}. And we have 
\begin{align}
	\boldsymbol{\alpha}_x(\vartheta_X)=\frac{1}{\sqrt{N_x}}\left[1, e^{\mathrm{i} \vartheta_X}, \cdots, e^{\mathrm{i} (N_x-1)\vartheta_X}\right],\nonumber\\
	\boldsymbol{\alpha}_y(\varphi_X)=\frac{1}{\sqrt{N_y}}\left[1, e^{\mathrm{i} \varphi_X}, \cdots, e^{\mathrm{i} (N_y-1)\varphi_X}\right],
\end{align}
Additionally, $\phi^\text{azi}_{G,i}$ ($\phi^\text{ele}_{G,i}$) and $\phi^\text{dep}_{G,i}$ denote the azimuth (elevation) angle of arrival (AOA) and the angle of departure (AOD) from the transmitter to the RISS (Tx-to-RISS). Furthermore, $\phi^\text{azi}_{h}$ ($\phi^\text{ele}_h$) represent the azimuth (elevation) angle of departure (AOD) from the RISS to the receiver (RISS-to-Rx), respectively, as depicts in the left-bottom of Fig. \ref{fig:systemmodel}. Thus, the received signal at the Rx can be formulated as 
\begin{align}
	y_\text{F} = \sum_{i=1}^2\sqrt{\varrho_{\text{Tx2R},i} \varrho_{\text{R2Rx}}}\mathbf{h}^T\boldsymbol{\Theta}\mathbf{G}_i\mathbf{v}_is_i+n_z,\label{eqn:farRec}
\end{align}
where $\varrho_{\text{Tx2R},i}$ and $\varrho_{\text{R2Rx}}$ denote the path loss between the Tx to the RISS and the RISS to the Rx, respectively. $\mathbf{v}_i, \forall i\in \{1,2\}$ denotes the transmit beamforming vector. $\boldsymbol{\Theta}=\text{diag}\{\phi_1,\cdots,\phi_N\}\in\mathbb{C}^{N\times N}$ denotes the reflection coefficient matrix of RISS and $|\phi_i|=1,\forall i\in N$ without loss of generality. $s_i,\forall i\in \{1,2\}$ is the $i^\text{th}$ normalized signal, i.e., $\mathbb{E}\{s_i^Hs_i\}=1$. $n_z\sim \mathcal{CN}(0, \sigma^2_0)$ is the additive Gaussian noise.
\subsection{Near-field Model}
We denote the location of the the $m^\text{th}$ antenna of the $i^\text{th}$ Tx as $\mathbf{b}_{\text{Tx},i,m}\in\mathbb{R}^{3\times 1},\forall i\in \{1,2\}, m\in M$ under a three-dimensional (3D) Cartesian coordinate system, while $\mathbf{b}_{\text{Rx}}\in\mathbb{R}^{3\times 1}$ and $\mathbf{b}_{\text{RISS},k}\in\mathbb{R}^{3\times 1},\forall k\in N$ denote the location of Rx and the location of $k^\text{th}$ element of the RISS, respectively. Furthermore, denote the distance between the $m^\text{th},m\in M$ antenna of the $i^\text{th}$ Tx and the $k^\text{th}$ element on the RISS as $d_{\text{Tx2R}}^{i,m,k},\forall i\in\{1,2\}, k\in N$, and the distance between the $k^\text{th}$ element and the Rx as $d_{\text{R2Rx}}^{k},\forall k\in N$. We have
\begin{align}
	&d_{\text{Tx2R}}^{i,m,k}=\left|\left|\mathbf{b}_{\text{Tx},i,m}-\mathbf{b}_{\text{RISS},k}\right|\right|,\label{eqn:dis1}\\
	&d_{\text{R2Rx}}^{k}=\left|\left|\mathbf{b}_{\text{RISS},k}-\mathbf{b}_{\text{Rx}}\right|\right|.\label{eqn:dis2}
\end{align}
Thus, the corresponding near-field channel \cite{nearfieldchannel} can be formulated as 
\begin{align}
	&\left[\bar{\mathbf{G}}_i\right]_{k,m} = \text{exp}\left(\frac{-\mathbbm{i}2\pi}{\lambda}d_{\text{Tx2R}}^{i,m,k}\right),i\in\{1,2\},k\in N,m\in M, \label{eqn:nearfG}\\
	&\left[\bar{\mathbf{h}}\right]_k = \text{exp}\left(\frac{-\mathbbm{i}2\pi}{\lambda}d_{\text{Tx2R}}^{k}\right),k\in N,\label{eqn:nearfh}
\end{align}
where $\bar{\mathbf{G}}_i\in\mathbb{C}^{N\times M}$ and $\bar{\mathbf{h}}\in\mathbb{C}^{N\times 1}$ denote the channel between the Tx to the RISS, and the RISS to the Rx, respectively. $\lambda$ denote the carrier wavelength. Similarly, we can derive the received signals at the Rx as 

\begin{align}
	y_{\text{N}}&=\sum_{i\in\{1,2\}}\sum_{m\in M}\sum_{k\in N}\sqrt{\varrho\left(d_{\text{Tx2R}}^{i,m,k}\right)\varrho\left(d_{\text{R2Rx}}^{k}\right)}\nonumber\\
	&\qquad\qquad\qquad\qquad\qquad \times \left[\bar{\mathbf{h}}\right]_k\phi_{k}\left[\bar{\mathbf{G}}_i\right]_{k,m}\left[\bar{\mathbf{v}}_i\right]_m+n_z\nonumber\\
	&\overset{(a)}{\approx} \sum_{i=1}^2\sqrt{\varrho_{\text{Tx2R},i} \varrho_{\text{R2Rx}}}\bar{\mathbf{h}}^T\boldsymbol{\Theta}\bar{\mathbf{G}}_i\bar{\mathbf{v}}_is_i+n_z,\label{eqn:nearRec}
\end{align}
where $(a)$ comes from the approximate that $\varrho\left(d_{\text{Tx2R}}^{i,m,k}\right)\approx \varrho_{\text{Tx2R},i},\forall m\in M, k\in N$ and $\varrho\left(d_{\text{R2Rx}}^k\right)\approx\varrho_{\text{R2Rx}}, \forall k\in N$.

Observe from Eq. \eqref{eqn:farRec} and Eq. \eqref{eqn:nearRec} that the far-field and near-field models share similar expressions\footnote{In practice, when the Tx and Rx are sufficiently far from the RIS, the waves transition from a near-field spherical wavefronts to far-field planar wavefronts, making the near-field and far-field models effectively equivalent. Generally, the far-field assumption holds when $R > \frac{2D^2}{\lambda}$. Here, $D$ represents the maximum dimension of the antenna, typically its diameter or maximum length. $R$ and $\lambda$ denote the distance between Tx/Rx and RIS and wavelength, respectively.}. However, the specific channel expressions (i.e., Eq. \eqref{eqn:farfG}-\eqref{eqn:farfh} and Eq. \eqref{eqn:nearfG}-\eqref{eqn:nearfh}) differ. We will compare the disparities between the two channels through theoretical algorithm design and practical examination in the subsequent sections\footnote{While our current study focuses on realizing interference suppression and signal enhancement through a convex optimization framework, the Fourier optics approach \cite{droulias2024reconfigurable} provides complementary insights by examining the RIS as a spatial filter through its transfer function.}.

\section{RISS-Aid Communications}\label{sec:III}
In this section, we demonstrate the potential of RISS in interference suppression scenarios. We consider a scenario where two transmitters simultaneously transmit communication signals, with one of the transmitters being undesired, denoted as Tx. 1, serving as the interference source, and Tx. 2 as the target source. The RISS employs active elements for sensing and guides the passive elements for reflection, effectively suppressing the interference signal while enhancing the signal-to-interference-noise ratio (SINR) at the Rx.
\subsection{Far-field RISS-Aid Communications}
With the assistance of RISS, we gain knowledge about the incident angles of both the interference signal and the target signal. Additionally, leveraging the capabilities of deep learning, signal incident identification can be recognized effectively, as demonstrated in \cite{CNNNetwork}. This recognition allows for the selective passage of signals by designing appropriate phase shift matrix for RISS. Specifically, the power of the incident signals can be reformulated as

\begin{align}
	&\varrho_{\text{Tx2R},i} \varrho_{\text{R2Rx}}\left|\mathbf{h}^T\boldsymbol{\Theta}\mathbf{G}_i\mathbf{v}_is_i\right|^2\nonumber\\
	=&\varrho_{\text{Tx2Rx},i}\left|\boldsymbol{\alpha}^T(\vartheta_{h}, \varphi_{h})\boldsymbol{\Theta}\boldsymbol{\alpha}(\vartheta_{G,i}, \varphi_{G,i})\boldsymbol{\beta}^T(\varpi_{G,i})\mathbf{v}_i\right|^2\nonumber\\
	&\qquad\qquad\qquad\qquad\qquad\qquad\qquad\qquad,\forall i\in \{1,2\}\label{eqn:calA}
\end{align}
where $\varrho_{\text{Tx2Rx},i}=\varrho_{\text{Tx2R},i} \varrho_{\text{R2Rx}}$.

\begin{Lemma}\label{lemma:1}
	Observe from Eq. \eqref{eqn:calA} that the optimal transmit beamforming for the $i^\text{th}$ transmitter can be derived as
\begin{align}
	&\mathbf{v}_i=\sqrt{P_i}\frac{\boldsymbol{\beta}^\dagger(\varpi_{G,i})}{||\boldsymbol{\beta}(\varpi_{G,i})||},\label{eqn:alignv}
\end{align}
where $P_i$ is the transmit power and $||\cdot||$ denotes the Euclidean norm. 
\end{Lemma}
\begin{IEEEproof}
	Note that Eq. \eqref{eqn:calA} can be decomposed into two sub-objectives, resulting in
	\begin{align}
		\underbrace{\boldsymbol{\alpha}^T(\vartheta_{h}, \varphi_{h})\boldsymbol{\Theta}\boldsymbol{\alpha}(\vartheta_{G,i}, \varphi_{G,i})}_{\text{The first objective}}\underbrace{\boldsymbol{\beta}^T(\varpi_{G,i})\mathbf{v}_i}_\text{The second objective}. \label{eqn:Myseg}
	\end{align}
	It is important to emphasize that our objective is to adjust the reflection matrix $\boldsymbol{\Theta}$ of the RISS through a sensing-assisted scheme, allowing the RISS to independently control the signals in free space. And from the Tx's perspective, the focus is solely on maximizing the term $\boldsymbol{\beta}^T(\varpi_{G,i})\mathbf{v}_i$.
	
	According to the Cauchy-Schwarz inequality, the maximum of $|\boldsymbol{\beta}^T(\varpi_{G,i}) \mathbf{v}_i|$ is achieved when $\mathbf{v}_i$ is aligned with the conjugate of $\boldsymbol{\beta}(\varpi_{G,i})$, i.e., $\mathbf{v}_i = \boldsymbol{\beta}^\dagger(\varpi_{G,i})$. This can then be normalized as Eq. \eqref{eqn:alignv}, which is consistent with the maximal ratio transmission (MRT) technique.
\end{IEEEproof}

\begin{Rem}
	The rationale behind this segmentation is to effectively decouple the transmitter and RISS. Specifically, the transmitter maximizes power in the RISS direction through beamforming, while the RISS utilizes its active elements for sensing and designs a reflection phase shift matrix to determine whether to enhance or suppress the signal. This design preserves the structure and design principles of traditional transceivers, ensuring spatial freedom for the signal at RISS. The proposed scheme enables the isolation of signals in the spatial domain at the RISS end, facilitating spatial filtering.
\end{Rem}

According to Lemma \ref{lemma:1}, Eq. \eqref{eqn:calA} can be rewritten as 
\begin{align}
	&\varrho_{\text{Tx2Rx},i}\left|\boldsymbol{\alpha}^T(\vartheta_{h}, \varphi_{h})\boldsymbol{\Theta}\boldsymbol{\alpha}(\vartheta_{G,i}, \varphi_{G,i})\boldsymbol{\beta}^T(\varpi_{G,i})\mathbf{v}_i\right|^2\nonumber\\
	&=\varrho_{\text{Tx2Rx},i}\left|\mathbf{c}_i\boldsymbol{\alpha}(\vartheta_{G,i},\varphi_{G,i})\right|^2\nonumber\\
	&=\varrho_{\text{Tx2Rx},i}\mathbf{c}_i\boldsymbol{\alpha}(\vartheta_{G,i},\varphi_{G,i})\boldsymbol{\alpha}^H(\vartheta_{G,i},\varphi_{G,i})\mathbf{c}_i^H\nonumber\\
	&=\varrho_{\text{Tx2Rx},i}\mathbf{c}_i\boldsymbol{\mathcal{A}}_i\mathbf{c}_i^H\nonumber\\
	&=\varrho_{\text{Tx2Rx},i}\text{trace}(\boldsymbol{\mathcal{C}}_i\boldsymbol{\mathcal{A}}),\forall i\in \{1,2\}\label{eqn:eq15}
\end{align}
where $\mathbf{c}_i=\sqrt{MP_i}\boldsymbol{\alpha}^T(\vartheta_{h},\varphi_{h})\boldsymbol{\Theta}$, $\boldsymbol{\mathcal{C}}_i=\mathbf{c}_i^H\mathbf{c}_i$, $\boldsymbol{\mathcal{A}}_i=\boldsymbol{\alpha}(\vartheta_{G,i},\varphi_{G,i})\boldsymbol{\alpha}^H(\vartheta_{G,i},\varphi_{G,i})$.

Thus, to maximize the target signal power while eliminate the interference signal power, the objective can be expressed as 
\begin{align}
		\text{(P1): }\max_{\boldsymbol{\Theta}} \quad&\text{trace}(\boldsymbol{\mathcal{C}}_2\boldsymbol{\mathcal{A}}_2)\label{eqn:eqnP1}\\
		\text{s.t.}\quad&\text{diag}(\boldsymbol{\mathcal{C}}_i)=MP_i\mathbf{1}_N,\forall i\in\{1,2\},\tag{\ref{eqn:eqnP1}a}\label{eqn:eqnP1consA}\\
		&\text{trace}(\boldsymbol{\mathcal{C}}_i)\leq MNP_i,\forall i\in \{1,2\},\tag{\ref{eqn:eqnP1}b}\label{eqn:eqnP1consB}\\
		&\text{trace}(\boldsymbol{\mathcal{C}}_1\boldsymbol{\mathcal{A}}_1)=0,\tag{\ref{eqn:eqnP1}c}\label{eqn:eqnP1consC}\\
		&\boldsymbol{\mathcal{C}}_i = \mathbf{c}_i^H\mathbf{c}_i\succeq \mathbf{0},\forall i\in \{1,2\},\tag{\ref{eqn:eqnP1}d}\label{eqn:eqnP1consD}\\
		&\mathbf{c}_i=\sqrt{MP_i}\boldsymbol{\alpha}^T(\vartheta_{h},\varphi_{h})\boldsymbol{\Theta}, \forall i\in \{1,2\},\tag{\ref{eqn:eqnP1}e}\label{eqn:eqnP1consE}\\
		&\text{Rank}(\boldsymbol{\mathcal{C}}_i) = 1,\forall i\in \{1,2\},\tag{\ref{eqn:eqnP1}f}\label{eqn:eqnP1consF}
\end{align}
where the objective $\text{(P1)}$ is to maximize the target power gain, with Eq. \eqref{eqn:eqnP1consA} to Eq. \eqref{eqn:eqnP1consC} derived from constraints imposed by the RISS passive phase shift, transmit power, and interference elimination. Here, $P_i$ represents the transmit power of the $i^\text{th}$ transmitter, and we set $P_i=P,\forall i\in \{1,2\}$ for simplification.

According to Lemma \ref{lemma:1} and $\text{(P1)}$, we can further simplify objective $\text{(P1)}$ as 

\begin{align}
	\text{(P1.1):  }\max_{\boldsymbol{\Theta}_G} \,\,\,&\text{trace}(\boldsymbol{\mathcal{C}}_2\boldsymbol{\mathcal{A}}_2)\label{eqn:eqnP1p1}\\
	\text{s.t.}\,\,\,\,\,&\mathbf{c}_i^T=\sqrt{MP}\mathbf{1}_N^T\boldsymbol{\Theta}_G,\tag{\ref{eqn:eqnP1p1}a}\label{eqn:eqnP1p1consA}\\
	&\text{trace}(\boldsymbol{\mathcal{C}}_1\boldsymbol{\mathcal{A}}_1)\leq \tau,\tag{\ref{eqn:eqnP1p1}b}\label{eqn:eqnP1p1consB}\\
	&\eqref{eqn:eqnP1consA}, \eqref{eqn:eqnP1consB},\eqref{eqn:eqnP1consD}, \tag{\ref{eqn:eqnP1p1}c}\label{eqn:eqnP1p1consC}
\end{align}
where \eqref{eqn:eqnP1p1consA} and \eqref{eqn:eqnP1p1consB} come from the further segmentation of Eq. \eqref{eqn:Myseg} as 
\begin{align}
	&\boldsymbol{\alpha}^T(\vartheta_{h}, \varphi_{h})\boldsymbol{\Theta}\boldsymbol{\alpha}(\vartheta_{G,i}, \varphi_{G,i})\nonumber\\
	=&\boldsymbol{\alpha}^T(\vartheta_{h}, \varphi_{h})\boldsymbol{\Theta}_h\boldsymbol{\Theta}_G\boldsymbol{\alpha}(\vartheta_{G,i}, \varphi_{G,i})\nonumber\\
	=&\mathbf{1}_N^T\boldsymbol{\Theta}_G\boldsymbol{\alpha}(\vartheta_{G,i}, \varphi_{G,i}),
\end{align}
where $\boldsymbol{\Theta}_h=\text{diag}\left(\boldsymbol{\alpha}^\dagger(\vartheta_{h}, \varphi_{h})\right)$. $\tau$ is a reasonable threshold to assess the extent of interference suppression.

Note that we have omitted the rank-one constraint in $\text{(P1.1)}$ compared to $\text{(P1)}$. Therefore, additional steps, such as Gaussian randomization or iterative rank minimization algorithms (IRM), are required for rank-one reconstruction. 

Inspired by \cite{IRMalg}, we utilize IRM algorithm for rank-one recovery. Specifically, we first obtain the solution of $\text{(P1.1)}$ and denote it as $\boldsymbol{\mathcal{C}}_2^0$. Once $\text{Rank}\left(\boldsymbol{\mathcal{C}}_2^0\right)=1$, the optimal solution $\boldsymbol{\Theta}=\boldsymbol{\Theta}_h\boldsymbol{\Theta}_G^0$ is achieved for the original objective $\text{(P1)}$. Otherwise, the objective $\text{(P1.1)}$ transforms into
\begin{align}
	\text{(P1.2):  }\min_{\boldsymbol{\Theta}_{G}^t} \,\,\,&-\text{trace}(\boldsymbol{\mathcal{C}}_{2}^t\boldsymbol{\mathcal{A}}_2)+\epsilon_t r \label{eqn:eqnP1p2}\\
	\text{s.t.}\,\,\,\,\,&\mathbf{c}_i^T=\sqrt{MP}\mathbf{1}_N^T\boldsymbol{\Theta}_G,\tag{\ref{eqn:eqnP1p2}a}\label{eqn:eqnP1p2consA}\\
	&\text{trace}(\boldsymbol{\mathcal{C}}_1\boldsymbol{\mathcal{A}}_1)\leq \tau,\tag{\ref{eqn:eqnP1p2}b}\label{eqn:eqnP1p2consB}\\
	&r\mathbf{I}_{N-1}-\mathbf{V}_{t}^H\boldsymbol{\mathcal{C}}_{2}^t\mathbf{V}_t \succeq \mathbf{0},\tag{\ref{eqn:eqnP1p2}c}\label{eqn:eqnP1p2consC}\\
	&\eqref{eqn:eqnP1consA}, \eqref{eqn:eqnP1consB},\eqref{eqn:eqnP1consD}, \tag{\ref{eqn:eqnP1p2}d}\label{eqn:eqnP1p2consD}
\end{align}
where $\mathbf{V}_{t}\in\mathbb{C}^{N\times N-1}$ denotes the matrix composed of eigenvectors corresponding to $[\lambda_1,\ldots,\lambda_{N-1}]$, and $\lambda_1<\lambda_2<\ldots<\lambda_{N}$ represent the ascending eigenvalues of $\boldsymbol{\mathcal{C}}_{2}^t$ in the $t^\text{th}$ iteration. $r$ is the positive relaxation variable. As $r$ approaches zero, the undesired eigenvalue $[\lambda_1,\cdots,\lambda_{N-1}]$ are forced to become zeros, which implies that $\boldsymbol{\mathcal{C}}_2^t$ satisfies the rank-one constraint. $\epsilon_l$ and $\varsigma$ represent gradually increasing auxiliary variables that control the rate at which $r$ decreases. Note that $\text{(P1.2)}$ is a convex optimization problem, which can be solved by standard optimization method, such as the interior-point method. The IRM algorithm based solution is summarized in Algorithm \ref{alg:1}.

\begin{algorithm}[t]
	\caption{IRM algorithm based solution for $\text{(P1.2)}$.}
	\begin{algorithmic}[1]\label{alg:1}
		\REQUIRE~\ $\boldsymbol{\mathcal{A}}$, $M$, $P$, $\tau$.
		\ENSURE~\ Optimal solution $\boldsymbol{\mathcal{C}}_2^t$ and optimal phase shift matrix $\boldsymbol{\Theta}$.
		\STATE Initialize $\epsilon_0=4$, $t=0$, $\varsigma=1.5$
		\STATE Obtain the preliminary solution $\boldsymbol{\Theta}_G^0$ from $\text{(P1.1)}$;
		\REPEAT
		\STATE Obtain the matrix $\mathbf{V}_t$ from eigenvalue decomposition;
		\STATE Solve the problem $\text{(P1.2)}$ and obtain $\boldsymbol{\Theta}_G^{t+1}$ and $r$;
		\STATE $\epsilon_{t+1} = \epsilon_{t}\varsigma$;
		\STATE $t=t+1$;
		\UNTIL{$r$ reaches a sufficiently small value or reaches the maximum iterations.}
		\STATE Obtain the optimal solution $\boldsymbol{\Theta}_{G}^t$ from eigenvalue decomposition of $\boldsymbol{\mathcal{C}}_2^t$.
		\STATE Obtain the optimal phase shift matrix by $\boldsymbol{\Theta}=\boldsymbol{\Theta}_{h}\boldsymbol{\Theta}_{G}^t$.

	\end{algorithmic}
\end{algorithm}

\subsection{Near-field RISS-Aid Communications}
Similar to the far-field model, we employ the RISS to enhance the target signal while suppressing the interference signal. Based on the near-field channel model, we can rewrite Eq. \eqref{eqn:nearRec} as 
\begin{align}
	y_{\text{N}}&\approx \sum_{i\in\{1,2\}}\sqrt{\varrho_{\text{Tx2R},i} \varrho_{\text{R2Rx}}}\sum_{m\in M}\sum_{k\in N}\left[\bar{\mathbf{h}}\right]_k\phi_{k}\left[\bar{\mathbf{G}}_i\right]_{k,m}\left[\bar{\mathbf{v}}_i\right]_m\nonumber\\
	&\qquad\qquad\qquad\qquad\qquad\qquad\qquad\qquad\qquad\qquad+n_z\nonumber\\
	&= \sum_{i\in\{1,2\}}\sqrt{\varrho_{\text{Tx2R},i} \varrho_{\text{R2Rx}}}\sum_{m\in M}\left[\bar{\mathbf{v}}_i\right]_m\nonumber\\
	&\times\sum_{k\in N}\text{exp}\left(\frac{-\mathbbm{i}2\pi}{\lambda}\left(d_{\text{Tx2R}}^{i,m,k}+d_{\text{R2Rx}}^{k}\right)+\mathbbm{i}\varphi_k\right)+n_z,
\end{align}
where $\phi_k=\text{exp}\left(\mathbbm{i}\varphi_k\right)$.

To direct attention towards the design of the reflection phase shift, we assume $M=1$ for simplification in the subsequent content. Thus, the target signal incident the Rx can be simplified as 

\begin{align}
	y_2 = &\sqrt{\varrho_{\text{Tx2R},2} \varrho_{\text{R2Rx}}}\nonumber\\
	&\quad\times\sum_{k\in N}\text{exp}\left(\frac{-\mathbbm{i}2\pi}{\lambda}\left(d_{\text{Tx2R}}^{2,k}+d_{\text{R2Rx}}^{k}\right)+\mathbbm{i}\varphi_k\right).
\end{align}

Represent $\left|y_2/\sqrt{\varrho_{\text{Tx2R},2} \varrho_{\text{R2Rx}}}\right|$ with $S_{2}$, we have 
\begin{align}
	S_2=\left|\sum_{k\in N}\text{exp}\left(\frac{-\mathbbm{i}2\pi}{\lambda}\left(d_{\text{Tx2R}}^{2,k}+d_{\text{R2Rx}}^{k}\right)+\mathbbm{i}\varphi_k\right)\right|.\label{eqn:eqnS2}
\end{align}
Similarly we have\footnote{Disregarding path loss, $S_1$ and $S_2$ represent the strengths of the interference signal and the target signal, respectively.}
\begin{align}
	S_1=\left|\sum_{k\in N}\text{exp}\left(\frac{-\mathbbm{i}2\pi}{\lambda}\left(d_{\text{Tx2R}}^{1,k}+d_{\text{R2Rx}}^{k}\right)+\mathbbm{i}\varphi_k\right)\right|.
\end{align}

Thus, we can formulate the objective as maximizing the reflected target signal while minimizing the interference signal through the design of the reflection phase shift matrix of the RISS, as
\begin{align}
	\text{(P2): }&\max_{\varphi_k,k\in N} \,S_2-\eta S_1\label{eqn:eqnP2}\\
	\text{s.t.}\,\,\,\,\,&S_i=\left|\sum_{k\in N}\text{exp}\left(\frac{-\mathbbm{i}2\pi}{\lambda}\left(d_{\text{Tx2R}}^{i,k}+d_{\text{R2Rx}}^{k}\right)+\mathbbm{i}\varphi_k\right)\right|\nonumber\\
	&\qquad\qquad\qquad\qquad\qquad\qquad\qquad, \forall i\in \{1,2\},\tag{\ref{eqn:eqnP2}a}\label{eqn:eqnP2consA}\\
	&\eta>0,\tag{\ref{eqn:eqnP2}b}\label{eqn:eqnP2consB}
\end{align}
where objective $\text{(P2)}$ aims to maximize the reflecting power of the target signal while minimizing the reflecting power of the interference signal. The coefficient $\eta$ is a constant representing the degree of interference suppression. It is noteworthy that $\text{(P2)}$ can be effectively solved by an alternating optimization (AO)-based algorithm, approximating the solution by optimizing one phase shift of the $N$ elements at a time while keeping the others constant. This process continues until the objective value in $\text{(P2)}$ converges.

In order to facilitate algorithm convergence, we set the initial phase shift matrix as $\eta=0$, indicating the sole maximization of the reflecting power of the target signal. Observing from Eq. \eqref{eqn:eqnS2}, the optimal $\varphi_k$ should satisfy
\begin{align}
	&\varphi_k-\frac{2\pi}{\lambda}\left(d_{\text{Tx2R}}^{2,k}+d_{\text{R2Rx}}^{k}\right)\nonumber\\
	&\qquad \qquad \qquad\qquad =\varphi_l-\frac{2\pi}{\lambda}\left(d_{\text{Tx2R}}^{2,l}+d_{\text{R2Rx}}^{l}\right)+ t2\pi,\nonumber\\
	&\qquad\qquad\qquad\qquad\forall k\in N, \forall l\in N, t=\pm1,\pm2,\cdots. \label{eqn:optimalPhi}
\end{align}
Note that Eq. \eqref{eqn:optimalPhi} ensures the effective stacking of each link from Tx. 2 to the Rx to maximize receiver power. We can further simplify the expression of Eq. \eqref{eqn:optimalPhi} as
\begin{align}
	\varphi_k=2\pi\left(\frac{d_{\text{Tx2R}}^{2,k}+d_{\text{R2Rx}}^{k}}{\lambda}-\left\lfloor\frac{d_{\text{Tx2R}}^{2,k}+d_{\text{R2Rx}}^{k}}{\lambda}\right\rfloor\right)+\psi ,\label{eqn:initialvalue}
\end{align}
where $\psi\in\left[0, 2\pi\right]$ denotes the flexible variable shared among all passive elements of the RISS. The AO-based algorithm is summarized in Algorithm \ref{alg:2}\footnote{Note that the input location information of Algorithm \ref{alg:2} can be directly estimated by the active elements of the RISS, which includes distance estimation \cite{distanceEST, distanceEST2} and DOA estimation\cite{MVDR,MUSIC,ESPRIT,ROOTMUSIC}. In this paper, we assume that the distances between the Tx, Rx, and the center of the RISS are fixed, while varying the angles of incident signals.}.

\begin{algorithm}[t]
	\caption{AO-based solution for $\text{(P2)}$.}
	\begin{algorithmic}[1]\label{alg:2}
		\REQUIRE~\ The location of Tx $\mathbf{b}_{\text{Tx},i,m}\in\mathbb{R}^{3\times 1},\forall i\in \{1,2\}, m\in M$, the location of the RISS $\mathbf{b}_{\text{RISS},k}\in\mathbb{R}^{3\times 1}$ and the Rx $\mathbf{b}_{\text{Rx}}\in\mathbb{R}^{3\times 1},\forall k\in N$, and wavelength $\lambda$.
		\ENSURE~\ Optimal passive phase shift matrix $\varphi_k,k\in N$.

		\STATE Calculate the distance $d_{\text{Tx2R}}^{i,k}$ and $d_{\text{R2Rx}}^{k}$ by Eq. \eqref{eqn:dis1} and Eq. \eqref{eqn:dis2};
		\STATE Obtain the initial phase shift by Eq. \eqref{eqn:initialvalue};
		\REPEAT 
		\FOR {k=1:N}
			\STATE Find the optimal $\varphi_k^*$ that maximizes the objective function (P2) while keeping the other $\{\varphi_n\}^N_{n=1,n\neq k}$ fixed;
			\STATE Update $\varphi_k=\varphi_k^*$;
		\ENDFOR
		\UNTIL{Converges or reaches the maximum iterations.}
	\end{algorithmic}
\end{algorithm}

\subsection{Robust Design} \label{sec:robustdesign}
\begin{figure}
	\centering
	\includegraphics[width=0.99\linewidth]{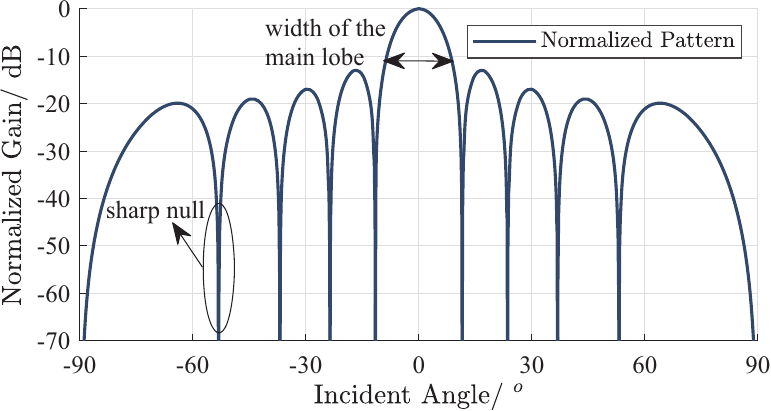}
	\caption{The reflecting normalized beam pattern, with the main-lobe directed at $0^\circ$.}
	\label{fig:sharpnull}
\end{figure}

To enhance the application potential of the RISS-assisted communication system, we consider sensing errors in this section. Specifically, due to the inherent width of the main-lobe, and the sharpness of nulls demonstrate in Fig. \ref{fig:sharpnull}, we aim to ensure the maximization of the target signal while simultaneously broadening the suppression of interference signals for robustness.

We assume that the Tx, Rx and the RISS are all on the same plane, and the location of the Rx is perfectly known (i.e., we have the knowledge of $\mathbf{b}_{\text{Rx}}$), while the actual incident angle of the interference signal (i.e., $\phi^\text{azi}_{G, 1}$) is within
\begin{align}
	\phi^\text{azi}_{G, 1}\in \left[\hat{\phi}{^\text{azi}_{G, 1}}-\delta, \hat{\phi}{^\text{azi}_{G,1}}+\delta\right],
\end{align}
where $\hat{\phi}{^\text{azi}_{G, 1}}$ denotes the estimated incident angle of Tx. 1, and $\delta>0$ represents the boundary of the DOA estimation error. The fundamental robust design approach for both far-field and near-field scenarios involves dividing the interval $[\hat{\phi}{^\text{azi}_{G, 1}}-\delta, \hat{\phi}{^\text{azi}_{G, 1}}+\delta]$ into $L$ sub-intervals with a step size of $2\delta/{(L-1)}$, denoted by $\hat{\phi}{^\S_{1,l}}, \forall l\in L$, and applying the suppression constraints to all angles within the interval.
\subsubsection{Robust Design for Far-field Model}
Specifically, to broaden the null of the reflecting beam pattern, we apply the constraints \eqref{eqn:eqnP1p1consB} to $\hat{\phi}{^\S_{1,l}}, \forall l\in L$, resulting in
\begin{align}
	\text{(P1.3):  }\min_{\boldsymbol{\Theta}_{G}^t} \,\,\,&-\text{trace}(\boldsymbol{\mathcal{C}}_{2}^t\boldsymbol{\mathcal{A}}_2)+\epsilon_t r \label{eqn:eqnP1p3}\\
	\text{s.t.}\,\,\,\,\,&\text{trace}(\boldsymbol{\mathcal{C}}_1\boldsymbol{\mathcal{A}}_{1,l})\leq \tau,\forall l\in L,\tag{\ref{eqn:eqnP1p3}a}\label{eqn:eqnP1p3consa}\\                
	&\eqref{eqn:eqnP1p1consA}, \eqref{eqn:eqnP1consA}, \eqref{eqn:eqnP1consB}, \eqref{eqn:eqnP1consD}, \tag{\ref{eqn:eqnP1p3}b}\label{eqn:eqnP1p3consB}
\end{align}
where $\boldsymbol{\mathcal{A}}_{1,l}=\boldsymbol{\alpha}(\hat{\vartheta}^\S_{1,l},\varphi_{G,1})\boldsymbol{\alpha}^H(\hat{\vartheta}^\S_{1,l},\varphi_{G,1})$, with $\hat{\vartheta}^\S_{1,l} = \pi\cos(\pi/2)=0$ due to the assumption that all the Tx, Rx, and the RISS are on the same plane. The IRM algorithm can be continuously used to solve the objective $\text{(P1.3)}$ for rank-one recovery.

\subsubsection{Robust Design for Near-field Model}
Similar to the far-field scenario, when the actual incident angle of the interference signal is within $\phi^\text{azi}_{G, 1}\in [\hat{\phi}{^\text{azi}_{G, 1}}-\delta, \hat{\phi}{^\text{azi}_{G,1}}+\delta]$, the location of Tx. 1 (i.e., $\mathbf{b}_{\text{Tx},1}$) and the distance between the $k^\text{th}$ RISS element and Tx. 1 vary, as depicted in Eq. \eqref{eqn:dis1}. Thus, the objective $\text{(P2)}$ can be rewritten as
\begin{align}
	\text{(P2.1):  }\max_{\varphi_k,k\in N} \,\,\,&S_2-\eta \max\{S_{1,l}\}\label{eqn:eqnP2p1}\\
	\text{s.t.}\,\,\,\,\,&d_{\text{Tx2R},l}^{1,k}=\left|\left|\mathbf{b}_{\text{Tx},1,l}-\mathbf{b}_{\text{RISS},k}\right|\right|,\tag{\ref{eqn:eqnP2p1}a}\label{eqn:eqnP2p1consA}\\
	&\eqref{eqn:eqnP2consA}, \eqref{eqn:eqnP2consB},\tag{\ref{eqn:eqnP2p1}b}\label{eqn:eqnP2p1consB}
\end{align}
where $S_{1,l}=\left|\sum_{k\in N}\text{exp}\left(\frac{-\mathbbm{i}2\pi}{\lambda}\left(d_{\text{Tx2R},l}^{1,k}+d_{\text{R2Rx}}^{k}\right)+\mathbbm{i}\varphi_k\right)\right|$. The detailed solution is outlined in Algorithm \ref{alg:2}.

\subsection{Computational Complexity and Convergence Analysis} \label{sec:complexityandconvergence}

This section analyzes the computational complexity of Algorithm \ref{alg:1} and Algorithm \ref{alg:1}. Specifically, the complexity of solving each optimization problem $\text{(P1.2)}$ using the interior-point method is $\mathcal{O}\left(N^{3.5} \log \frac{1}{\zeta}\right)$, where $\zeta$ denotes the prescribed accuracy \cite{luo2010semidefinite}. Let $I_{\mathrm{IRM}}$ denote the number of iterations required by the IRM algorithm, where each iteration involves an eigenvalue decomposition, incurring a computational cost of $\mathcal{O}(N^3)$. Consequently, the total computational complexity of Algorithm \ref{alg:1} can be expressed as $\mathcal{O}\left(I_{\mathrm{IRM}} \left(N^{3.5} \log \frac{1}{\zeta} + N^3\right)\right).$

In contrast, solving the optimization problem $\text{(P2)}$ requires an exhaustive search over $\Delta_\varphi N$ phase shift combinations, where each search involves recalculating $S_2 - \eta S_1$, introducing a complexity of $\mathcal{O}(N + \Delta_\phi N)$. Here, $\Delta_\varphi$ and $\Delta_\phi$ represent the phase shift resolution and the robust design resolution, respectively. Thus, the total computational complexity of Algorithm \ref{alg:2} with $I_{\mathrm{AO}}$ iterations is approximately $\mathcal{O}\left(I_{\mathrm{AO}} \Delta_\varphi N \left(N + \Delta_\phi N\right)\right) \approx \mathcal{O}\left(I_{\mathrm{AO}} \Delta_\varphi \Delta_\phi N^2\right).$

\begin{figure}
	\centering
	\subfloat[Convergence of Algorithm 1.]{\includegraphics[width=0.85\linewidth]{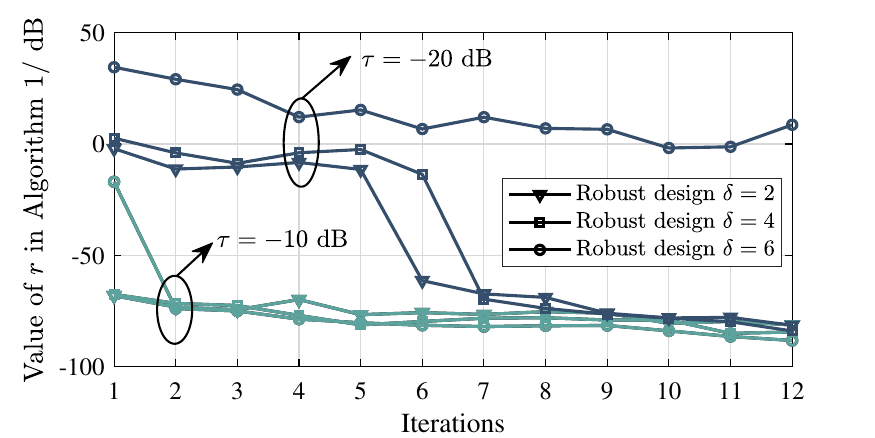}}
	\hfill
	\subfloat[Convergence of Algorithm 2.]{\includegraphics[width=0.85\linewidth]{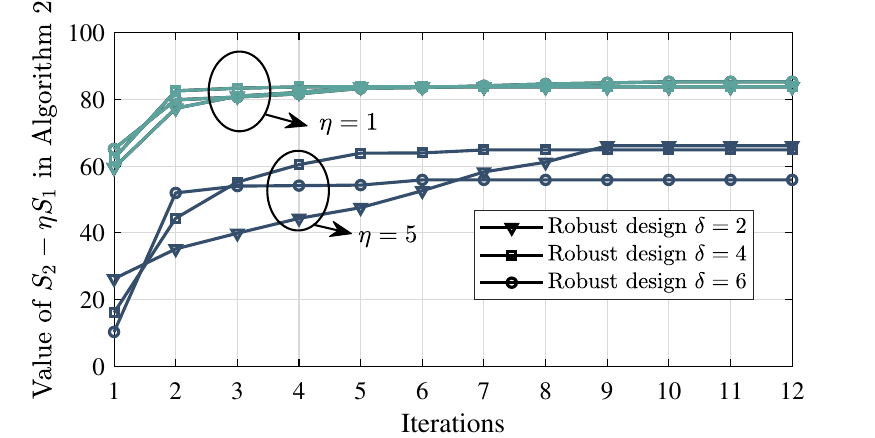}}
	\caption{Convergence of both algorithm, where $N=100$, $\Delta_\varphi=4$ for 2-bit RISS, and $\Delta_\phi=2\delta/0.1$.}
	\label{fig:conver}
\end{figure}

The detailed convergence analyses are available in Appendix \ref{app:A}, and the convergence behavior of both algorithms is illustrated in Fig. \ref{fig:conver}. Observed that both algorithms converge within a finite number of iterations. Furthermore, the number of required iterations increases when the interference suppression constraints become more stringent, e.g., $\tau = -20$ dB and $\eta = 5$. Therefore, appropriate interference suppression strategies can be selected based on the practical scenario to balance the trade-off between computational efficiency and system performance.

\section{Hardware Introductions and System Implements}\label{sec:hardwareintro}\label{sec:IV}
In this section, we introduce the actual system hardware of the RISS-assisted system, which includes two single horn antenna transmitters (i.e., $M=1$) and one single horn antenna receiver. Moreover, the RISS consists of $N_a=4$ active elements for sensing and $N=100$ passive elements for reflecting.
\subsection{Hardware Introductions}\label{sec:hardwareintro_A}
\begin{figure*}
	\centering
	\begin{tikzpicture} 
		\scope[nodes={inner sep=0,outer sep=0}]
		\node[anchor=south east] (a)
		{\includegraphics[width=8.5cm]{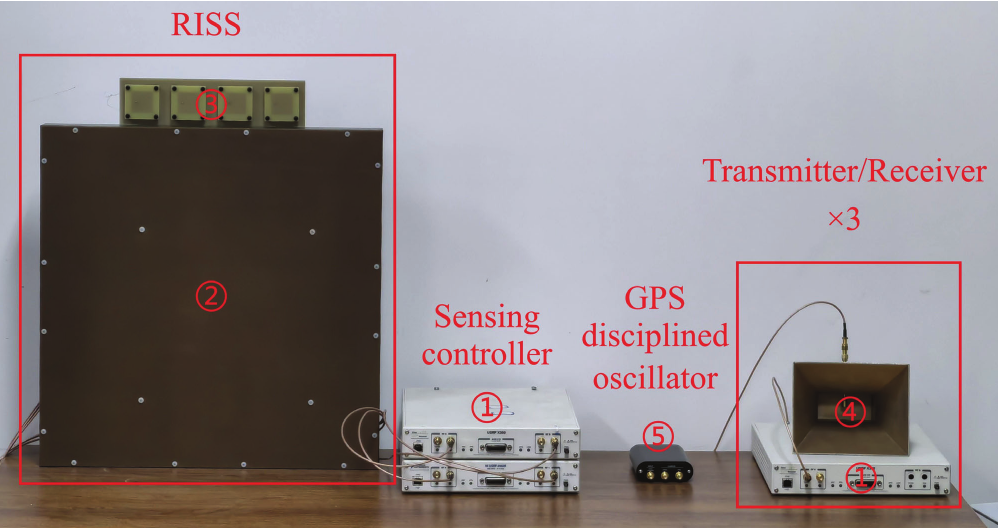}};

		\node[anchor=north east] (b)
		{\includegraphics[width=8.49cm]{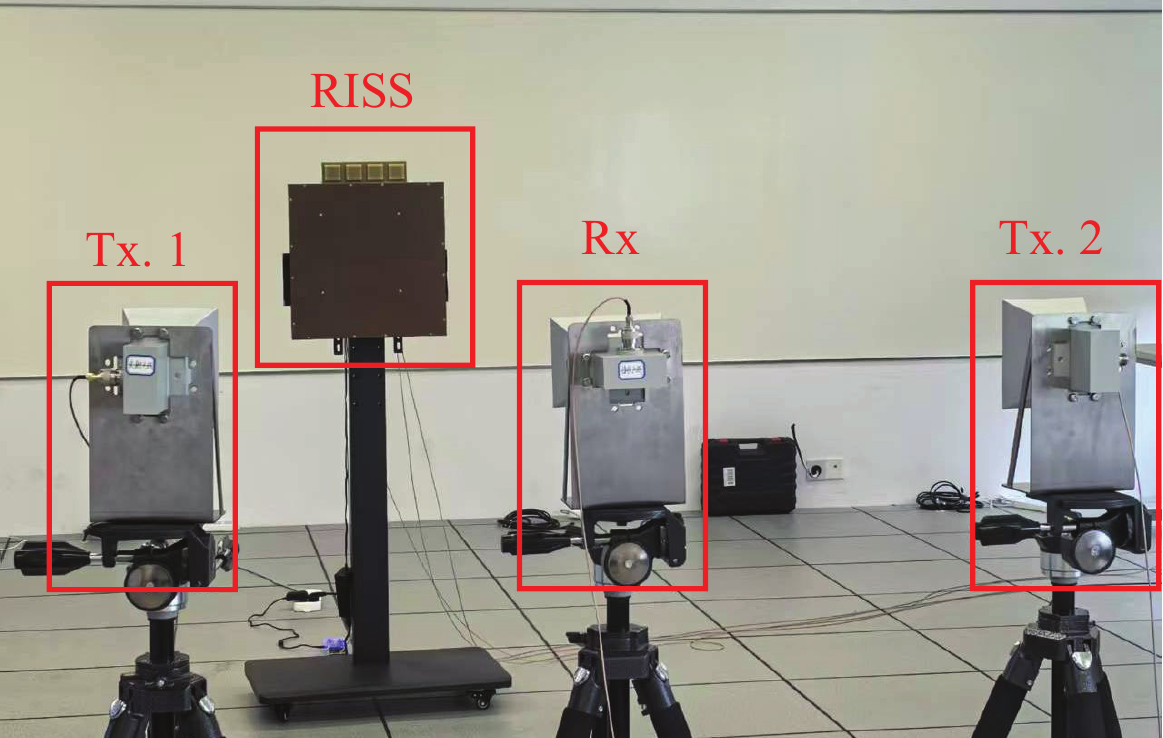}};

		\node[right=0.5mm of a.north east, anchor=north west] (c)
		{\includegraphics[height=9.85cm]{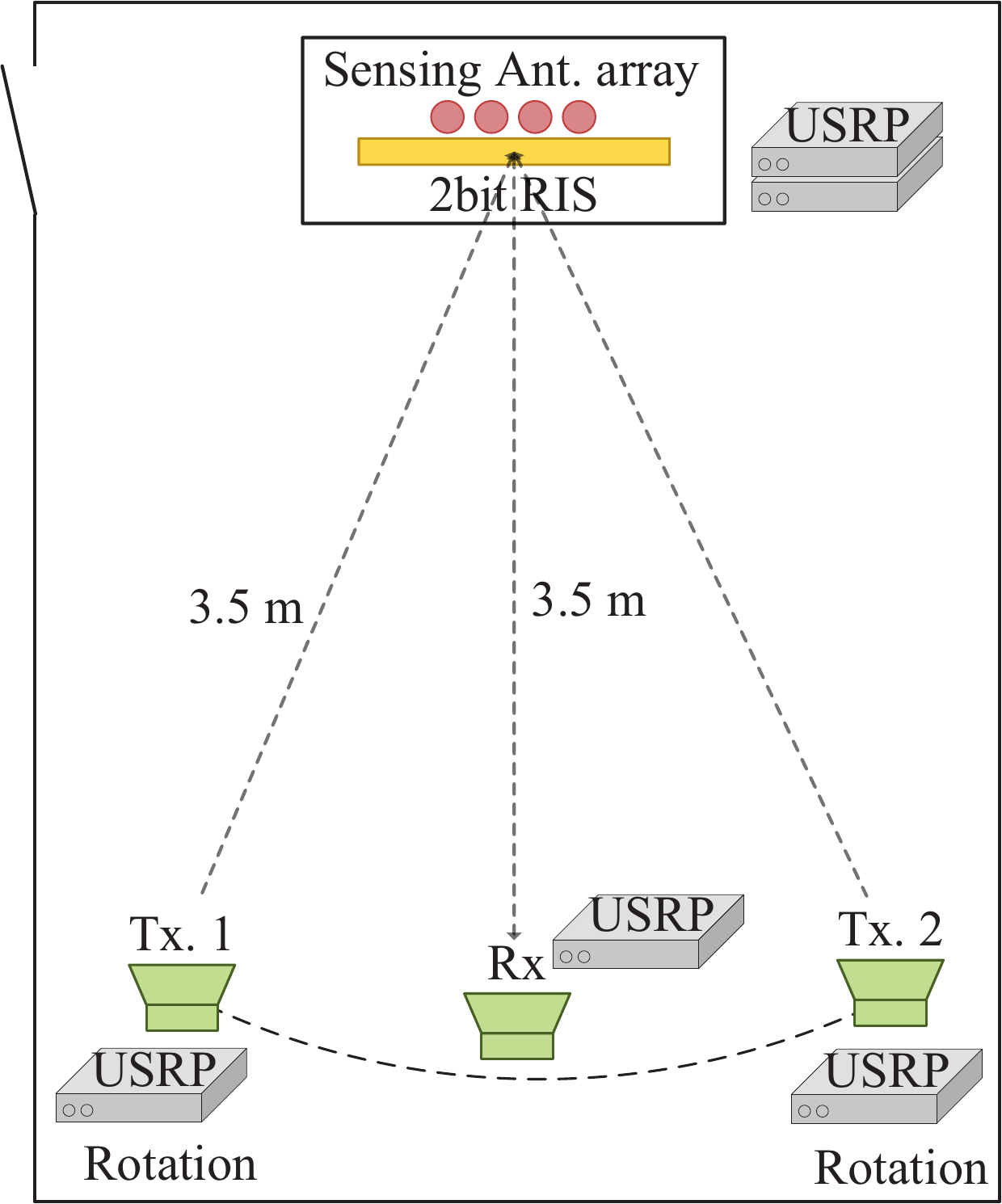}};

		\endscope
		\foreach \n in {a,b,c} {
		\node[anchor=north west,fill=green] at (\n.north west){(\n)};
		}
	\end{tikzpicture}
	\caption{Experiment measurement setup. $(a)$ The hardware list. $(b)$ Platform units arrangement. $(c)$ Laboratory floor plan.}
	\label{fig:hardwareintro}
\end{figure*}

This section introduces the hardware employed within the system and delineates its implementation throughout. Fig. \ref{fig:hardwareintro} presents an overview of the system's hardware, while Table-\ref{table:relatedworks} elaborates on the detailed parameters and functions of these hardware components. Note that we utilize an additional external clock (i.e., GPSDO) to synchronize local frequencies across different USRPs. Additional hardware parameters and descriptions, including the detailed specifications of the 2-bit RIS, the three-dimensional structural model, the $S_{11}$ reflection coefficient curve, and the radiation pattern of the sensing antenna array, are available in Appendix \ref{app:B}, which may offer further insights.

The hardware of entire system can be divided into three components, including communication component, sensing component and reflecting component. 
\subsubsection{Communication hardware} The communication hardware is facilitated by two transmitters and a single receiver. Each transmitter and receiver consists of a USRP paired with a horn antenna, enabling communication via QPSK modulation at $f=3.5$ GHz.
\subsubsection{Sensing hardware} The sensing hardware consists of two cascaded USRPs and form a $N_a=4$ RF link linear sensing array receiver. 
\subsubsection{Reflecting hardware} The reflecting hardware consists of a single RISS, which involves $N=100$, $N_a=4$ and operate at $f=3.5$ GHz.

\begin{table*}[t]
    \centering
    \caption{Hardware list of the entire system.}
    \begin{tabular}{m{1 cm}<{\centering}|m{2.9 cm}<{\centering}|m{5.0 cm}<{\centering}|m{5.8 cm}<{\centering}}
    \hline
    \textbf{Num.}&\textbf{Hardware} &  \textbf{Size}&  \textbf{Parameters and functions}\\
    \hline
    \ding{192}&Software Defined Radios (SDR)&/&
	\begin{itemize}
	\setlength{\itemindent}{-5mm}
	\item USRP X300/X310 series.
	\item Each USRP equipped with two UBX-160/SBX-120 series daughter-boards.
	\vspace{-0.3cm}
	\end{itemize}\\
    \hline
    \ding{193} &Reconfigurable Intelligent Surface (RIS)&430 mm (W) $\times$ 430 mm (H) $\times$ 46 mm (L) & 
	\begin{itemize}
		\setlength{\itemindent}{-5mm}
		\item 2-bits phase adjustable.
		\item $10\times 10$ elements.
		\item Frequency band $3.4\sim3.6$ GHz.
		\item Element spacing 42.83 mm, i.e., $0.5\lambda$, where $\lambda$ is the wavelength at 3.5GHz.
		\vspace{-0.3cm}
	\end{itemize}\\
	\hline
	\ding{194} &Sensing Antenna Array& 236 mm(W) $\times$ 59 mm (H) $\times$ 9.2 mm (L) & 
	\begin{itemize}
		\setlength{\itemindent}{-5mm}
		\item Four elements linear array microstrip antenna.
		\item Frequency band $3.08\sim3.91$ GHz.
		\item Total gain 12.6 dBi.
		\item Element spacing 58.96 mm, i.e., $0.6883\lambda$, where $\lambda$ is the wavelength at 3.5GHz.
		\vspace{-0.3cm}
	\end{itemize}\\
	\hline
	\ding{195} &Horn Antennas& 144 mm(W) $\times$ 114 mm (H) $\times$ 175 mm (L) & 
	\begin{itemize}
		\setlength{\itemindent}{-5mm}
		\item 10 dB gain horn antenna.
		\item Frequency band $2.60\sim3.95$ GHz.
		\vspace{-0.3cm}
	\end{itemize}\\
	\hline
	\ding{196} &GPS Disciplined Oscillator (GPSDO)& 70 mm(W) $\times$ 24 mm (H) $\times$ 130 mm (L) & 
	\begin{itemize}
		\setlength{\itemindent}{-5mm}
		\item 10MHz SC-Cut High Stability Oven Controlled Crystal Oscillator (OCXO).
		\vspace{-0.3cm}
	\end{itemize}\\
	\hline

    \end{tabular} \label{table:relatedworks}
\end{table*}

\subsection{Communication Model} \label{sec:hardwareintro_B}
\begin{figure}
	\centering
	\includegraphics[width=0.8\linewidth]{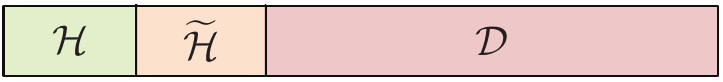}
	\caption{Frame structure of transmitted QPSK signals, where $\mathcal{H}$, $\tilde{\mathcal{H}}$ and $\mathcal{D}$ denote the synchronize head frame, the interval frame and data frame, respectively.}
	\label{fig:framestruc}
\end{figure}

In this system, we construct two transmitters simultaneously transmitting QPSK signals at 3.5 GHz to the receiver, simulating a scenario with interference present. Our goal is to isolate and receive only one of these signals, treating the signal from the other transmitter as interference. Fig. \ref{fig:framestruc} illustrates the signal's frame structures. It is evident from Fig. \ref{fig:framestruc} that the synchronization head frame serves the purpose of conducting autocorrelation operations at the receiver to determine the start of a frame. To maintain the synchronization head frame's consistency, we incorporate an additional interval frame, separating it from the data frame\footnote{Due to the upsampling and pulse-shaping filter process, the synchronization head frame changes with the data frame. The interval frame ensures a fixed synchronization head independent of the data frame, making it convenient for signal processing and recognition.}, resulting in a fixed synchronize head frame and simplify the system. To effectively distinguish between the two transmitted signals, we assign distinct values to the synchronization head frames of each transmitter signal, denoted as $\mathcal{H}_1=[1,0,1,0,1,1,0,1,0,1,...]$ and $\mathcal{H}_2=[1,0,0,0,1,1,0,0,0,1,...]$, while $\tilde{\mathcal{H}_i}=\sim\mathcal{H}_i, \forall i=\{1,2\}$, without loss of generality. And the detailed parameters are summarized in Table-\ref{table:communicationpara}. Note that $|\mathcal{D}|$ denotes the length of $\mathcal{D}$, and 5000 bits per frame means that we can transmit 2500 QPSK symbols per frame. $R_b$ and $f_s$ denote the symbol rate and sample rate, respectively. 

\begin{table}[t]
    \centering
	\caption{The frame structure and parameters of communication model}
	\begin{tabular}{c|c|c|c|c|c|c}
    \hline
	\multirow{2}*{$|\mathcal{D}|$}  & \multirow{2}*{$|\mathcal{H}|$}& \multirow{2}*{$R_b$}& \multirow{2}*{$f_s$}&\multicolumn{3}{c}{Raised cosine filter}\\ 
	\cline{ 5 - 7 }
	& & & & sps& span& $\beta$\\
	\hline
	5000 bits& 500 bits& 100 KHz & 1.6 MHz & 16 & 16 & 0.15\\ 
	\hline
	\end{tabular}\label{table:communicationpara}
\end{table}

\subsection{Sensing Model}\label{sec:hardwareintro_C}
We emphasize again that our primary aim is to selectively receive the target signal from the mixed signal through the assistance of sensing. The sensing model comprises two tasks: DOA estimation and the identification recognition of the two transmitters.
\subsubsection{DOA Estimation} The DOA estimation algorithms based on the linear array
have been extensively researched, e.g., Minimum Variance Distortionless Response (MVDR) \cite{MVDR}, Multiple Signal Classification (MUSIC) \cite{MUSIC}, Estimation of Signal Parameters using Rotational Invariance Techniques (ESPRIT) \cite{ESPRIT}, and their variations also have been studied and developed widely. We opt for MUSIC/ROOT-MUSIC algorithm as our DOA estimation scheme after conducting our actual measurement, and we further assume that both transmitter, receiver and RISS are on the same plane. Specifically, while the two signals incident from $\phi_\text{Tx2R}^1$ and $\phi_\text{Tx2R}^2$, the received signals at the sensing antenna array can be formulated as 
\begin{align}
	\mathbf{x}(n) = \mathbf{A}\mathbf{s}(n)+\mathbf{n}_z(n),
\end{align}
where $\mathbf{x}(n)\in \mathbf{C} ^{N_a\times 1}$ is the received signals from sensing array at snapshot $n$, $\mathbf{A}=\left[\boldsymbol{\alpha}\left(\phi_{\text{Tx2R}}^1\right), \boldsymbol{\alpha}\left(\phi_\text{Tx2R}^2\right)\right]\in\mathbb{C}^{N_a\times 2}$ is the direction matrix, and $\boldsymbol{\alpha}\left(\phi_{\text{Tx2R}}^i\right)=\left[1, e^{-\mathbbm{i}\phi_\text{Tx2R}^i},\cdots, e^{-\mathbbm{i}(N_a-1)\phi_\text{Tx2R}^i}\right]^T, \forall i\in \{1,2\}$ is the direction vectors. $\mathbf{s}(n)\in\mathbb{C}^{2\times 1}$ and $\mathbf{n}_z(n)\in\mathbb{C}^{N_a\times 1}$ denote the transmit signals and zero-mean additive white Gaussian noise (AWGN) vector at snapshot $n$ with the normalized noise power $\sigma_0^2$.

Assuming that the two transmitters are statistically independent of each other, thus
\begin{align}
	\mathbb{E}\left\{s_k(n) s_i^\dagger(n)\right\}= \begin{cases}P_k, & k=i \\ 0, & k \neq i\end{cases},\forall i,k\in\{1,2\}.
\end{align}

The covariance matrix of $\mathbf{x}\left(n\right)$ is given by
\begin{align}
	\mathbf{R}=\mathbb{E}\left\{\mathbf{x}(n) \mathbf{x}^{H}(n)\right\}=\mathbf{A P A}^{H}+\sigma_0^2 \mathbf{I}_{N_a},\label{eqn:musicR}
\end{align}
where $\mathbf{P}=\mathbb{E}\left\{\mathbf{s}(n)\mathbf{s}^\dagger(n)\right\}=\text{diag}\{P_1, P_2\}$, and $P_i,\forall i\in\{1, 2\}$ denotes the average power of the $i$-th signal. 

Following the procedures of the MUSIC algorithm, the eigenvalue decomposition of the matrix $\mathbf{R}$ in Equation \eqref{eqn:musicR} is initially derived as 
\begin{align}
	\mathbf{R}=
	\left[\mathbf{E}_s, \mathbf{E}_z\right]
	\left[
		\begin{array}{ll}
		\boldsymbol{\Xi}_s & \\
		& \boldsymbol{\Xi}_z
		\end{array}
	\right]
		\left[\begin{array}{c}
		\mathbf{E}_s^H \\
		\mathbf{E}_z^H
		\end{array}\right],
\end{align}
where $\mathbf{E}_s\in\mathbb{C}^{N_a\times 2}$ and $\mathbf{E}_z\in\mathbb{C}^{N_a\times (N_a-2)}$ denote the signal and noise subspaces, respectively, spanned by their corresponding eigenvectors. $\boldsymbol{\Xi}_s$ and $\boldsymbol{\Xi}_z$ denote the corresponding eigenvalues. Due to the orthogonality between $\boldsymbol{\alpha}\left(\phi_{\text{Tx2R}}^i\right),\forall i \in\{1,2\}$ and $\mathbf{E}_z$, the pseudo spectrum of MUSIC algorithm \footnote{The ROOT MUSIC \cite{ROOTMUSIC} algorithm follows a process similar to that of the MUSIC algorithm, with the key distinction lying in the final step, where roots of the equation are sought instead of employing a spectral search as shown in Eq. \eqref{eqn:spectrumSearch}.} can be formulated as 
\begin{align}
	P_{\mathrm{MUSIC}}(\phi)=\frac{1}{\boldsymbol{\alpha}(\phi)^H \mathbf{E}_z \mathbf{E}_z^H \boldsymbol{\alpha}(\phi)}, \phi\in[-\pi,\pi]. \label{eqn:spectrumSearch}
\end{align}

\subsubsection{Signal Recognition}
The utilization of signal recognition algorithms plays a pivotal role in identifying incident signals and bolstering network security\cite{signalrecognition1, signalrecognition2}. In this section, we present a deep-learning-based algorithm, ultimately augmenting the performance of the RISS-assisted system. Specifically, we harness the capabilities of a deep-learning-based algorithm to identify synchronization head frames. This allows us to make informed decisions on whether to enhance or suppress the incident signals\footnote{RF fingerprint recognition using deep learning has a rich history spanning decades, enabling the extraction of features attributable to hardware defects in transmitter circuits, such as frequency offset and phase shift. This capability facilitates the identification of transmitters. To ensure the system's feasibility, we choose to leverage more prominent features, i.e., the synchronization heads $\mathcal{H}_i,\forall i\in \{1,2\}$, for the purpose of classification.}.

\begin{figure}
	\centering
	\includegraphics[width=0.99\linewidth]{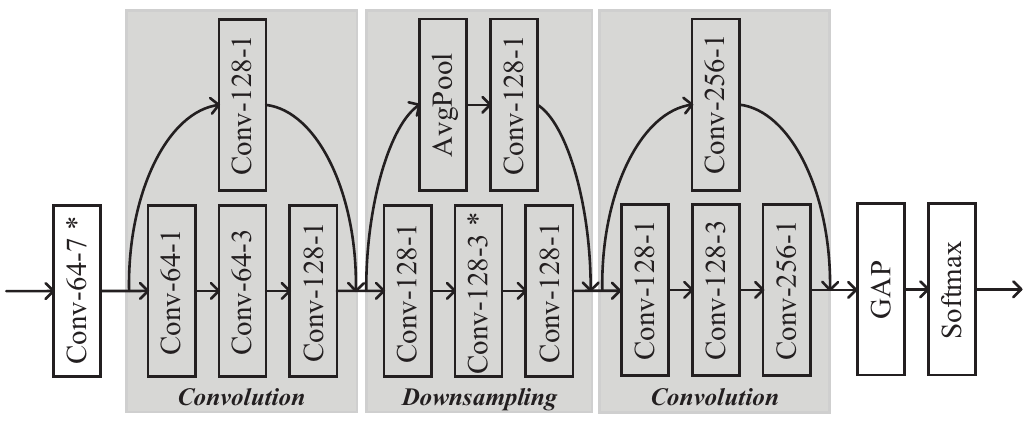}
	\caption{The structure of SRnet. The layer labeled "Conv-k-s" signifies that this convolutional layer incorporates $k$ kernels, each of size $s$. All convolutions with a size of 3 are separable. The asterisk $(*)$ indicates that this layer employs a stride of size 2. Batch normalization (BN) is implemented subsequent to each convolution, followed by activation using rectified linear units (ReLUs).}
	\label{fig:recogNet}
\end{figure}

Fig. \ref{fig:recogNet} illustrates the network architecture we employ for signal recognition, referred to as the signal recognition network (SRnet). The network begins with an initial convolutional layer, followed by a sequence of alternating convolution and downsampling blocks, and concludes with a classification layer. The initial convolutional layer consists of 64 kernels of size 7 with a stride of 2. Both the convolution and downsampling blocks are implemented as residual blocks, inspired primarily by ResNet-B and ResNet-D, as described in \cite{he2019bag}. The final classification layer performs global average pooling (GAP) before making predictions via the softmax function. The number of residual blocks can be adjusted to achieve varying levels of modeling capacity. The ResNet used in this study comprises two consecutive convolution blocks, with a downsampling block serving as a transitional stage.

\begin{figure}
	\centering
	\includegraphics[width=0.85\linewidth]{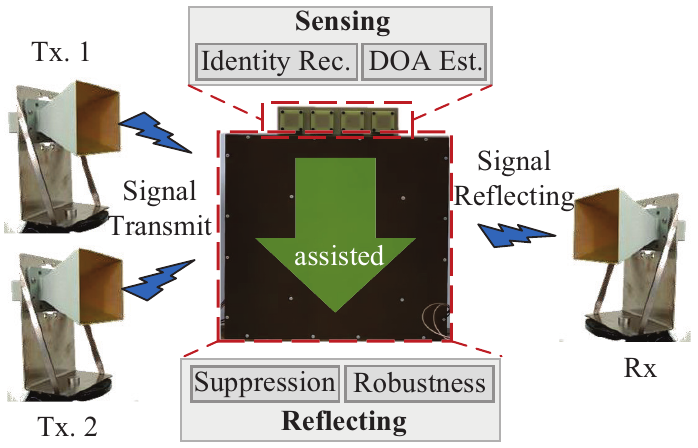}
	\caption{The entire process of the RISS-assisted system. 
	}
	\label{fig:entireprocess}
\end{figure}

The entire process of the RISS-assisted system is depicted as Fig. \ref{fig:entireprocess}. Specifically, when the signal is incident on the active elements, DOA information and identity information are recognized and utilized to design the reflected beamforming. The signal is then reflected to the Rx through the passive elements.

\section{Experiment Results}\label{sec:V}
In this section, we practical verify the performance of RISS-assisted system. The entire system including two single horn-antenna Tx (i.e., $M=1$) and one single horn-antenna Rx, which are both controlled by the USRP X300 and transmit/receive carrier QPSK signals, as demonstrated in Section \ref{sec:hardwareintro_A} to \ref{sec:hardwareintro_B}. The location of the center of the RISS is set to $[0, 0, 0]$. The distance between Tx/Rx and the center of the RISS is set to $d_c=3.5$ meters, and the Tx/Rx and RISS are on the same plane, which means that $\phi^\text{ele}=\pi/2$, thus, the location of the Tx and Rx is set to $\mathbf{b}_{\text{Tx}, i}=[d_c\sin(\phi^\text{azi}_i), 0, d_c\cos(\phi^\text{azi}_i)], \forall i\in\{1,2\}$ and $\mathbf{b}_{\text{Rx}}=[0, 0, d_c]$. In the experiments, we can obtain $\phi^\text{azi}_i$ from the DOA algorithm, e.g., MUSIC and ROOTMUSIC, and the identification can be also obtained and determine which Tx is the target source, as depicts in Section \ref{sec:hardwareintro_C}. The interference suppression factors are set to $\tau = 0.1$ (i.e., $-10$ dB) and $\eta=1$, respectively\footnote{Note that excessively large values of $\tau$ and $\eta$ may lead to a greater emphasis on interference suppression, potentially at the expense of target signal enhancement. Therefore, when selecting the values of both parameters, it is essential to strike a balance between interference suppression and the enhancement of the target signal, in accordance with the specific requirements of the system.}. Moreover, the Error Vector Magnitude (EVM) is adopted as a key metric to assess the modulation accuracy in digital communication systems. Typically expressed as a percentage, EVM quantifies the deviation between the received signal $s_{received}$ and the ideal reference signal $s_{ideal}$. It is mathematically defined as
\begin{align}
\Delta_\text{EVM} (\%) = \frac{RMS(s_{received}-s_{ideal})}{RMS(s_{ideal})} \times 100\%,
\end{align}
where RMS$(\cdot)$ denotes the root mean square operation. The RMS values for the error vector and the ideal signal are given by
\begin{align}
	&RMS(s_{received}-s_{ideal}) = \sqrt{\frac{1}{d} \sum_{d=1}^{|\mathcal{D}|/2} \left|s_{received, d} - s_{ideal, d}\right|^2}, \nonumber\\
	&RMS(s_{ideal}) = \sqrt{\frac{1}{d} \sum_{d=1}^{|\mathcal{D}|/2} \left|s_{ideal, d}\right|^2},
\end{align}
where $|\mathcal{D}|/2$ denotes the number of modulated symbols, given that the $|\mathcal{D}|$ bits are modulated using QPSK with a modulation order of 2.

\subsection{Sensing Performance}
\begin{figure}
	\centering
	\includegraphics[width=0.85\linewidth]{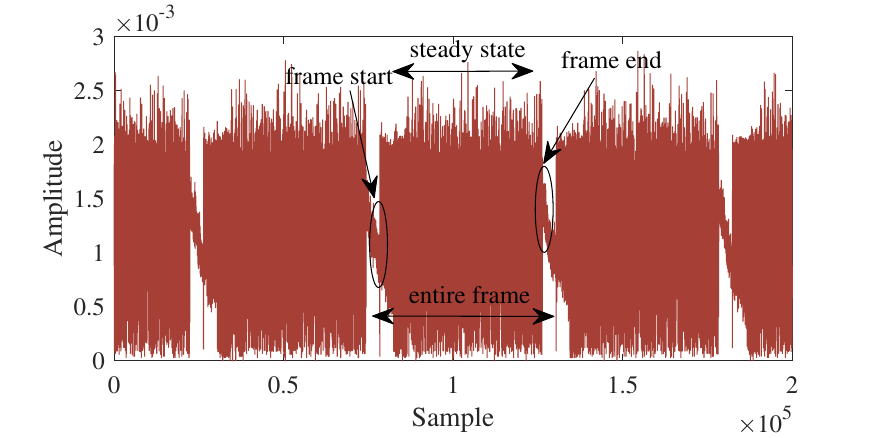}
	\caption{The amplitude of the samples within a block of USRP sampling varies across the different phases of a frame, which includes the start state, steady state, and end state.}
	\label{fig:block_data}
\end{figure}

As the foundation of the RISS-assisted system, we first validate the effectiveness of the sensing algorithm. Due to the hardware characteristics, we collect 200000 samples in each USRP sampling block, with each block containing multiple frames. Each frame comprises 48000 samples, including 2,500 information symbols and 500 synchronization head frames, with 16 samples per symbol. For the sake of stability, we preprocess the signal within a USRP sampling block and extract the steady-state segments for DOA estimation, as illustrated in Fig. \ref{fig:block_data}. The snapshot size for MUSIC/ROOTMUSIC DOA estimation is set to 5000 samples.

\begin{figure}
	\centering
	\includegraphics[width=0.85\linewidth]{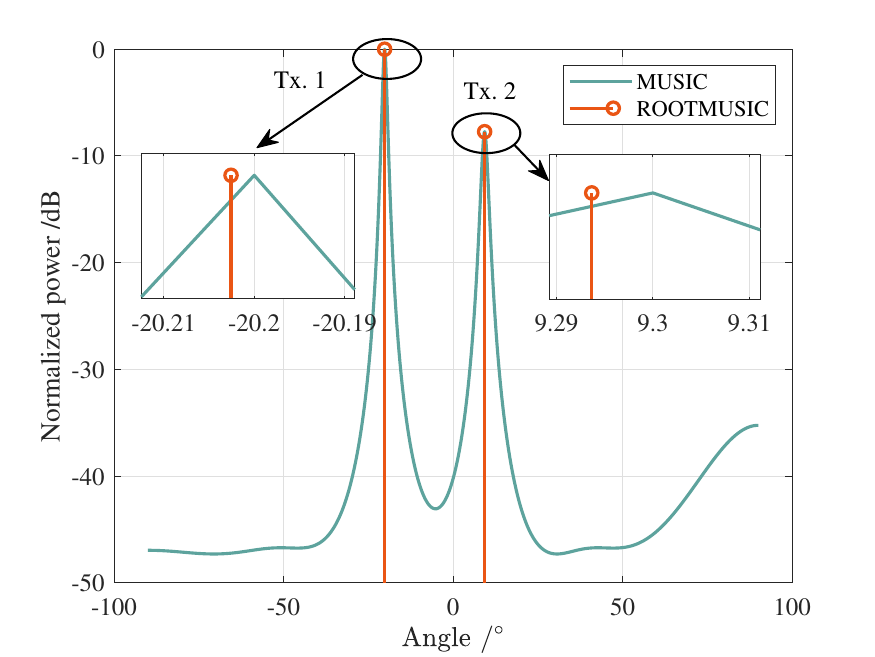}
	\caption{The experiment results of both MUSIC and ROOTMUSIC implemented by the sensing antenna array within the RISS. The Tx. 1 is located in $\phi^\text{azi}_1=-20^\circ$ and Tx. 2 is located in $\phi^\text{azi}_2=10^\circ$.   
	}
	\label{fig:DOA_performance}
\end{figure}

Fig. \ref{fig:DOA_performance} demonstrates that both the MUSIC and ROOTMUSIC algorithms can effectively estimate the two incident signals with slight error, thereby enabling more in-depth experimentation. 

Table-\ref{table:identificationRcog_para} presents parameters and performance of the proposed SR-net. The transmit gains of the two USRPs are set to 8 dB and 5 dB, respectively, corresponding to actual transmit powers of -19.46 dBm and -21.63 dBm. A total of 100 sampling blocks are collected and randomly divided into training, validation, and test sets in a 3:1:1 ratio.

Prior to inputting signal samples into SR-net, several pre-processing steps are required:
\begin{enumerate}
\item DOA estimation: This crucial step involves determining the directions of the incident signals, including both the target and interference signals.
\item Spatial filtering: This process is used to separate signals from the combined received data, resulting in distinct samples for the target and interference signals. Specifically, the samples are weighted and summed individually to implement spatial filtering based on the DOA results, thereby distinguishing and obtaining two separate sets of signal samples from the spatial characteristics. For further details, please refer to the MVDR \cite{MVDR} and Linearly Constrained Minimum Variance (LCMV)\cite{LCMV} methods.
\end{enumerate}

\begin{figure}
	\centering
	\includegraphics[width=0.75\linewidth]{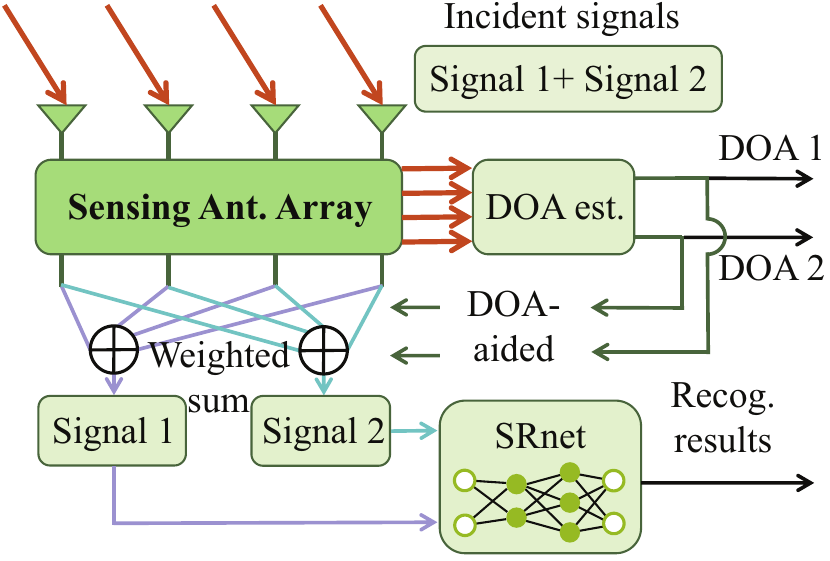}
	\caption{The processes of DOA estimation, spatial filtering and RF fingerprint recognition.   
	}
	\label{fig:processFlaw}
\end{figure}

Following the pre-processing steps, two signal sets are obtained and used for binary classification with SR-net. The SR-net is implemented using PyTorch \cite{pytorch2019} and trained on a single NVIDIA RTX A6000 GPU, utilizing the Adam optimizer \cite{kingma2014adam} for 100 epochs. After training the model, it is deployed offline on the upper computer connected to the RISS for sensing. The detialed processes are depicted in Fig. \ref{fig:processFlaw}.

\begin{table}[t]
    \centering
	\caption{The parameters and performance of SR-net.}
	\begin{tabular}{c|c|c|c|c}
	\hline
	Dev. & Tx Gain &Actual Tx Power & Blocks & Acc.\\
    \hline
	Tx. 1& 8 dB  & -19.46 dBm& \multirow{2}*{100}&\multirow{2}*{99.3\%}\\ 
	\cline{ 1 - 3 }
	Tx. 2& 5 dB & -21.63 dBm &\\
	\hline
	\end{tabular}\label{table:identificationRcog_para}
\end{table}

\subsection{Sensing-assisted Communication for Signal Enhancement}
\begin{figure}[!t]
	\centering
	\subfloat[RISS off, $\Delta_{\text{EVM}}=44.7494\%$.]{\includegraphics[width=0.48\linewidth]{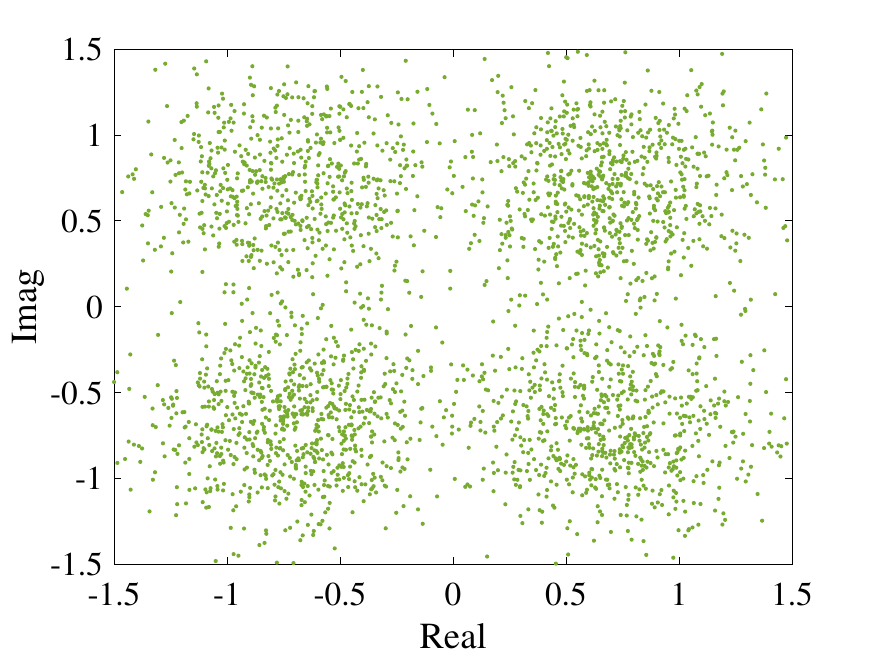}
	\label{fig:RISSOFF}}
	\hfill
	\subfloat[RISS on, $\Delta_\text{EVM}=26.4598\%$. ]{\includegraphics[width=0.48\linewidth]{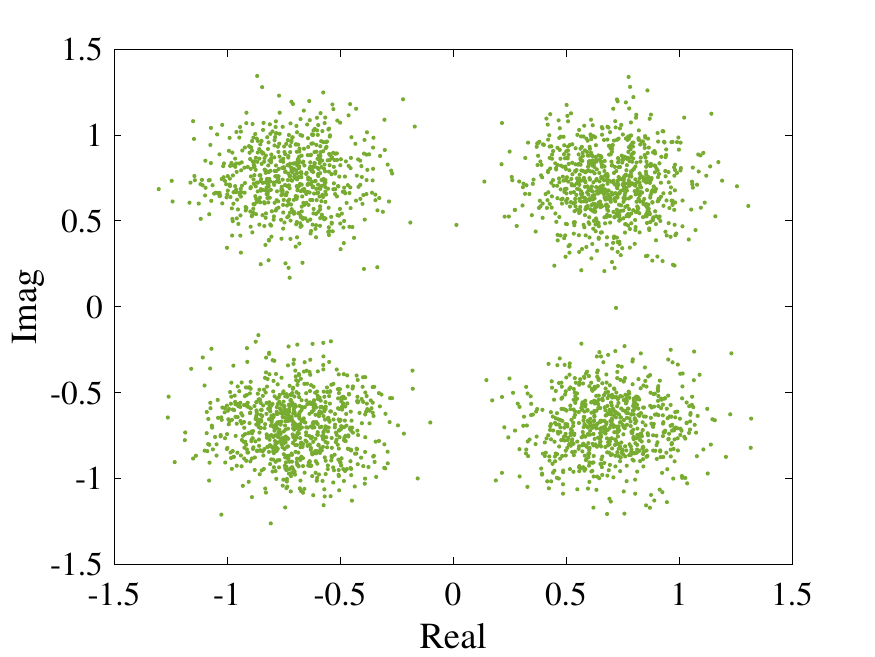}
	\label{fig:RISSON}}
	\caption{Constellation diagram measurements of the Rx. $(a)$ Turn off the RISS. $(b)$ Turn on the RISS, but all the phase shifts are set to $0^\circ$.}
	\label{fig:constellation_ONOFF}
\end{figure}

In this section, we first shows the sensing-assisted communication with only one Tx, i.e., $\tau=+\infty$ in Eq. \eqref{eqn:eqnP1p3} and $\eta = 0$ in Eq. \eqref{eqn:eqnP2p1} for simplification. This means that our sole objective in this section is to enhance the target signal without considering interference suppression\footnote{For detailed implementation videos, please refer to \url{https://drive.google.com/file/d/11PmI_luWSJ0AxLBtrOmpnz8PKYNMoFP4/view?usp=drive_link}.}. Fig. \ref{fig:constellation_ONOFF} demonstrates that RISS enhance the communication performance although the phase shifts of the RISS are not appropriate, since the RISS can increase multipath and change the polarization direction of the signal\footnote{The RISS reverses the polarization, resulting in the polarization of the Tx and Rx being opposite.}.


\begin{figure}
	\centering
	\subfloat[Near-field model with perfect sensing, $\Delta_\text{EVM}=4.1357\%$.]{\includegraphics[width=0.48\linewidth]{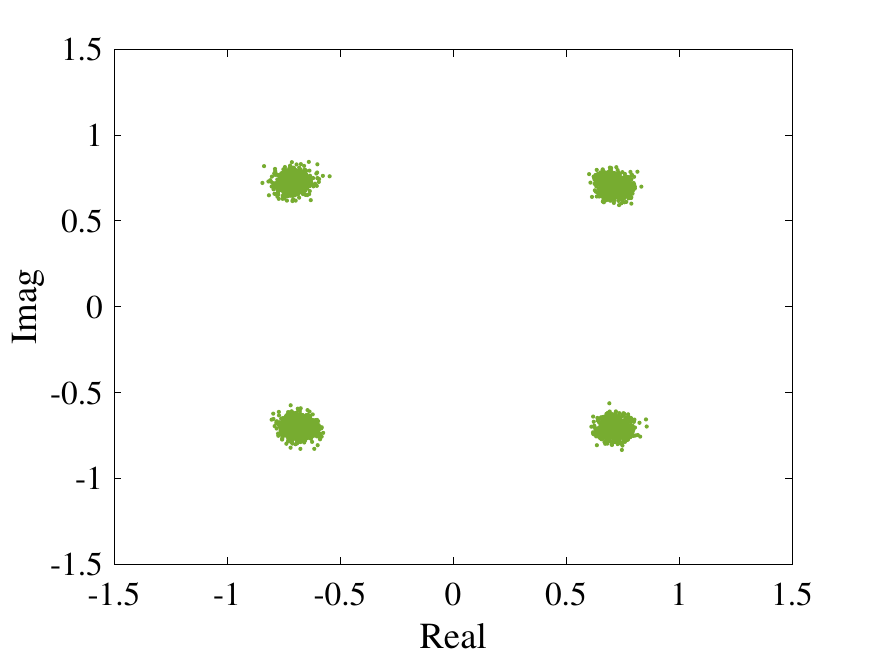}\label{fig:NF_perfect}}
	\hfill
	\subfloat[Near-field model with imperfect sensing, $\Delta_\text{EVM}=4.6768\%$.]{\includegraphics[width=0.48\linewidth]{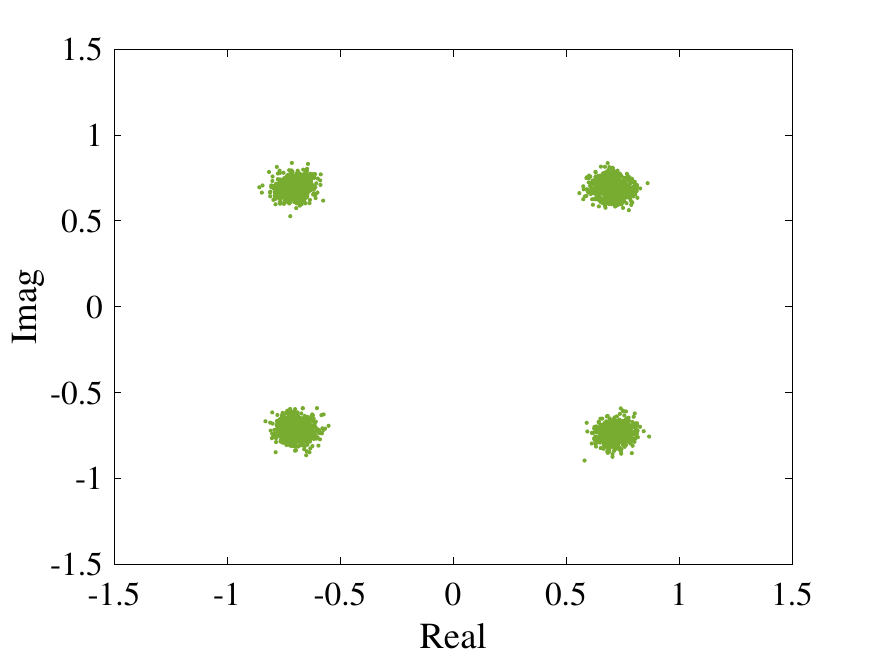}\label{fig:NF_imperfect}}
	\newline
	\subfloat[Far-field model with perfect sensing, $\Delta_\text{EVM}=5.1486\%$.]{\includegraphics[width=0.48\linewidth]{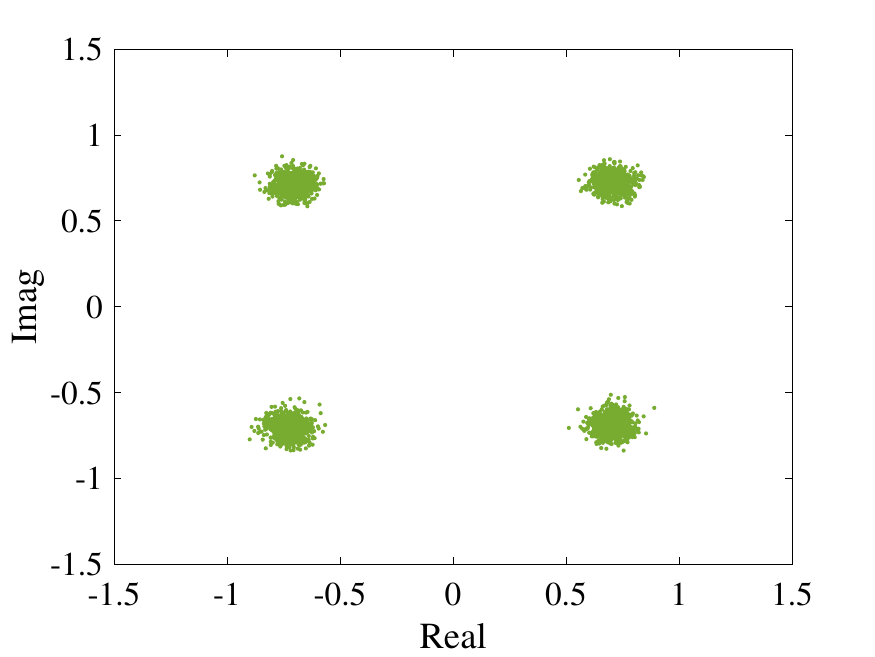}\label{fig:FF_perfect}}
	\hfill
	\subfloat[Far-field model with imperfect sensing, $\Delta_\text{EVM}=5.2401\%$.]{\includegraphics[width=0.48\linewidth]{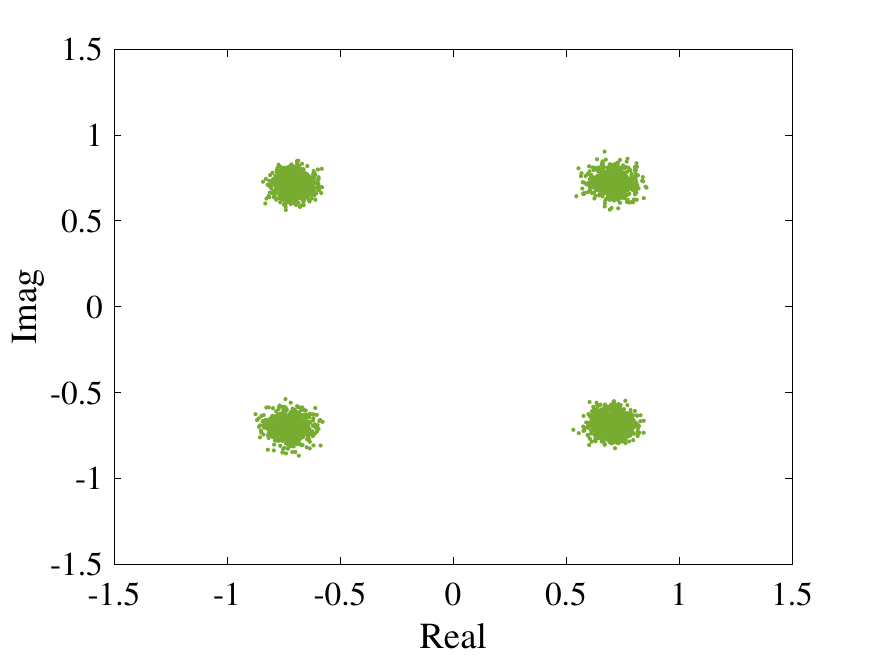}\label{fig:FF_imperfect}}

	\caption{Constellation diagram measurements of the Rx. $(a)$ Near-field model with perfect sensing. $(b)$ Near-field model with imperfect sensing. $(c)$ Far-field model with perfect sensing. $(d)$ Far-field model with imperfect sensing.}
	\label{fig:constellation_onlyalign}
\end{figure}

Fig. \ref{fig:constellation_onlyalign} demonstrates the constellation diagram measurements of the receiver. Note that perfect sensing assumes knowledge of the direction of Tx. 2, i.e., $\phi_2^\text{azi}=10^\circ$. In contrast, imperfect sensing refers to cases where the angle is derived from actual sensing results. As observed in Fig. \ref{fig:NF_perfect} and \ref{fig:FF_perfect}, both near-field and far-field models enhance communication performance and significantly improve the SINR at the receiver compared to Fig. \ref{fig:RISSON}. This improvement enables the adoption of higher-order modulation schemes, such as 16QAM and 64QAM, which enhance communication efficiency.

\begin{figure}
	\centering
	\includegraphics[width=0.85\linewidth]{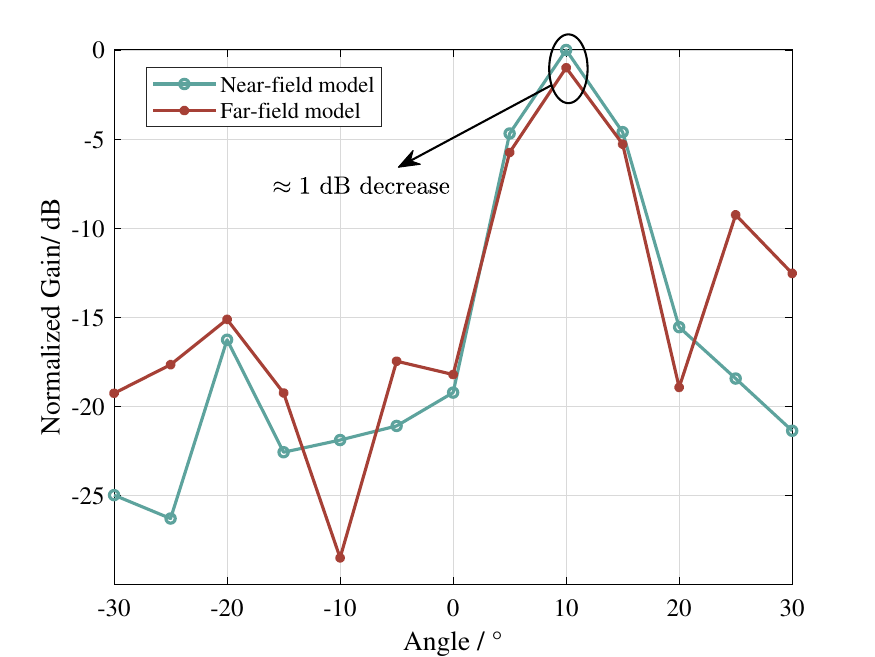}
	\caption{The normalized beampattern of near-field model and far-field model. The objective of both scheme is to maximize the signal power at the Tx with the knowledge of $\phi_2^\text{azi}=10^\circ$.}
	\label{fig:NF_FF_beampattern}
\end{figure}

Fig. \ref{fig:NF_imperfect} and Fig. \ref{fig:FF_imperfect} further illustrate that sensing errors have a slight impact on the communication performance of the target signal. This is because the main-lobe possesses an inherent width, as demonstrated in Fig. \ref{fig:sharpnull} and discussed in Section \ref{sec:robustdesign}. To further illustrate the phenomenon, the normalized beampattern measurements are presented in Fig. \ref{fig:NF_FF_beampattern}. Observed from that the main-lobe exhibits robustness for the target signal. Additionally, a side-lobe at $-20^\circ$ is evident in both the near-field and far-field models, which motivates the exploration of robust interference suppression schemes in the following section.

\subsection{Sensing-assisted Communication for Interference Suppression}\label{sec:V_c}
\begin{figure}[!t]
	\centering
	\subfloat[Only align the target signal, $\Delta_\text{EVM}=19.2229\%$.]{\includegraphics[width=0.48\linewidth]{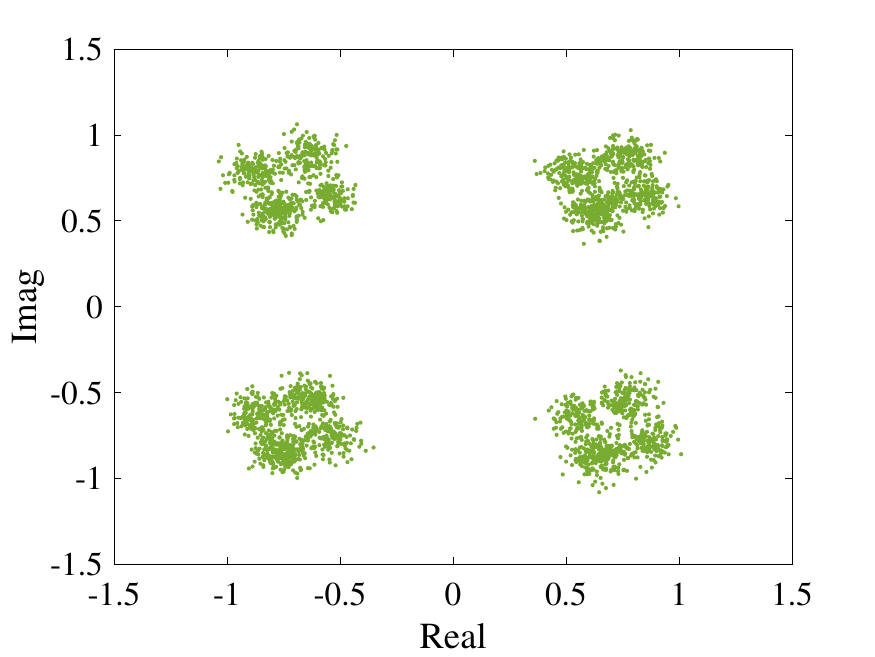}
	\label{fig:RISS_align}}
	\hfill
	\subfloat[Robust interference suppression, $\Delta_\text{EVM}=6.5897\%$.]{\includegraphics[width=0.48\linewidth]{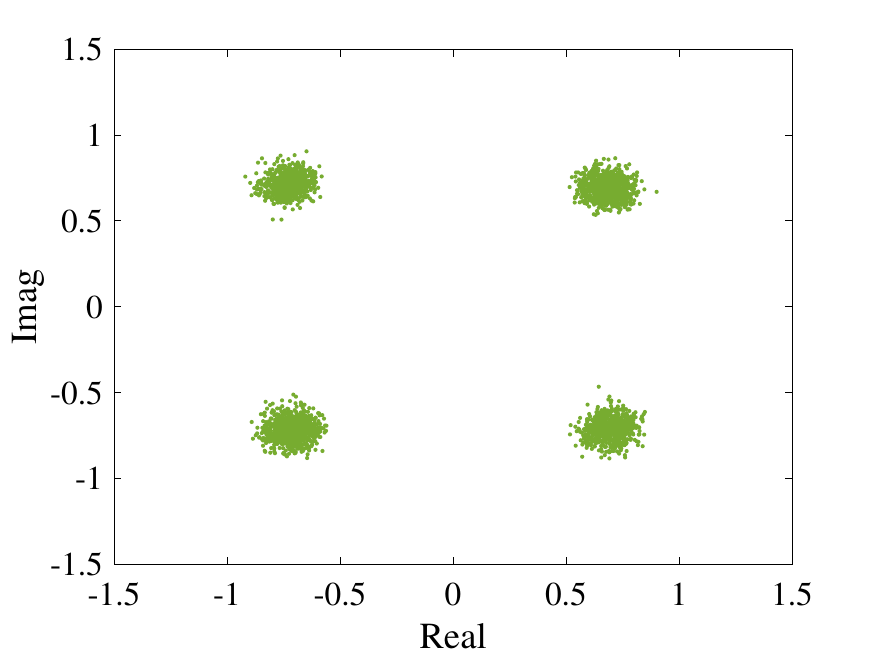}
	\label{fig:RISSinterSupp}}
	\caption{Constellation diagram measurements of the Rx in near-field model.}
	\label{fig:constellation_align_supp}
\end{figure}

\begin{figure}
	\centering
	\subfloat[Theoretical beam gain with 2-bit RISS.]{\includegraphics[width=0.85\linewidth]{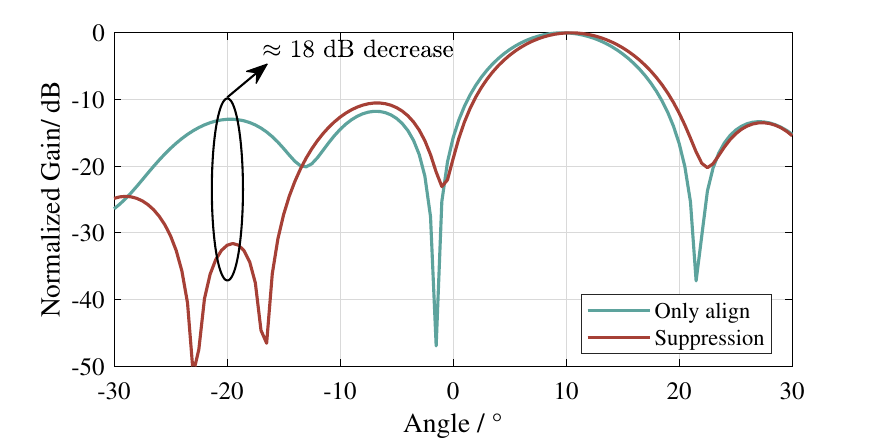}\label{fig:Beam_NF_theo}}
	\hfill
	\subfloat[Practical experimental beam gain.]{\includegraphics[width=0.85\linewidth]{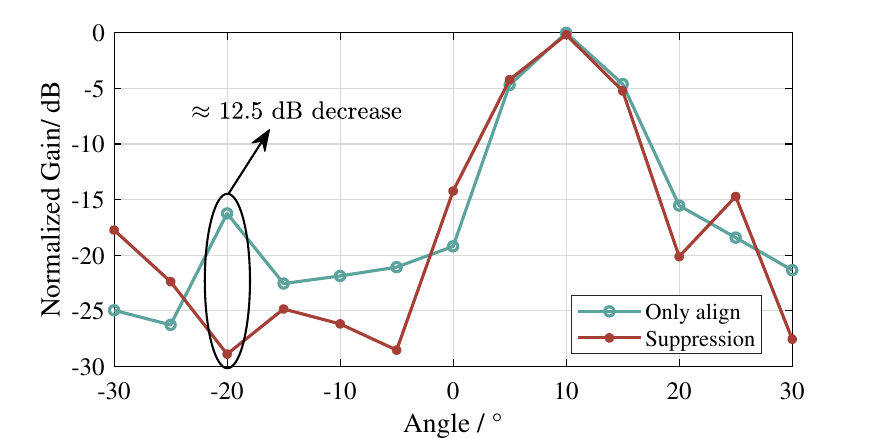}\label{fig:Beam_NF_prac}}
	\caption{The beampattern of align scheme and interference suppression scheme in the near-field model. The target signal is located in $\phi_2^\text{azi}=10^\circ$, while the interference signal is located in $\phi_1^\text{azi}=-20^\circ$.}
	\label{fig:align_suppre_beampattern}
\end{figure}

In this section, we further evaluate the performance of interference suppression\footnote{For detailed implementation videos, please refer to \url{https://drive.google.com/file/d/11SJT2nxz9Zuo5t7ZtBZR4MCcfASvaUpO/view?usp=drive_link}.}. The directions of the target signal and interference signal are set to $\phi^\text{azi}_2=10^\circ$ and $\phi^\text{azi}_1=-20^\circ$, respectively. As shown in Fig. \ref{fig:NF_FF_beampattern}, the interference signal is located in the side-lobe direction, leading to strong interference at the receiver. Fig. \ref{fig:constellation_align_supp} presents the communication constellation diagram with interference suppression using the near-field model, compared to the counterpart scheme that only aligns with the target signal. It can be observed that the interference signal reduces the SINR and causes the constellation diagram to become more dispersed. Furthermore, the constellation diagram exhibits a pattern similar to that of non-orthogonal multiple access (NOMA), due to the power disparity between Tx. 1 and Tx. 2. However, the interference suppression scheme significantly improves the SINR, restoring the compactness of the constellation diagram. The theoretical and experimental beampatterns of both schemes are shown in Fig. \ref{fig:align_suppre_beampattern}. It is noteworthy that both the theoretical and experimental results presented in Fig. \ref{fig:Beam_NF_theo} and Fig. \ref{fig:Beam_NF_prac} demonstrate a consistent trend in interference suppression and target signal enhancement achieved through the 2-bit RISS. Specifically, a gain is observed in the direction of the target signal, while deep nulls are formed in the directions of interference. However, due to practical system imperfections such as phase shift deviations in the reflection elements and environmental disturbances, the actual level of interference suppression exhibits a noticeable deviation from the theoretical prediction. Additionally, it is also observed that the interference suppression scheme results in a slight reduction in the main-lobe gain compared to the only align scheme.

\begin{figure}[!t]
	\centering
	\subfloat[Only align the target signal, $\Delta_\text{EVM}=43.5608\%$.]{\includegraphics[width=0.48\linewidth]{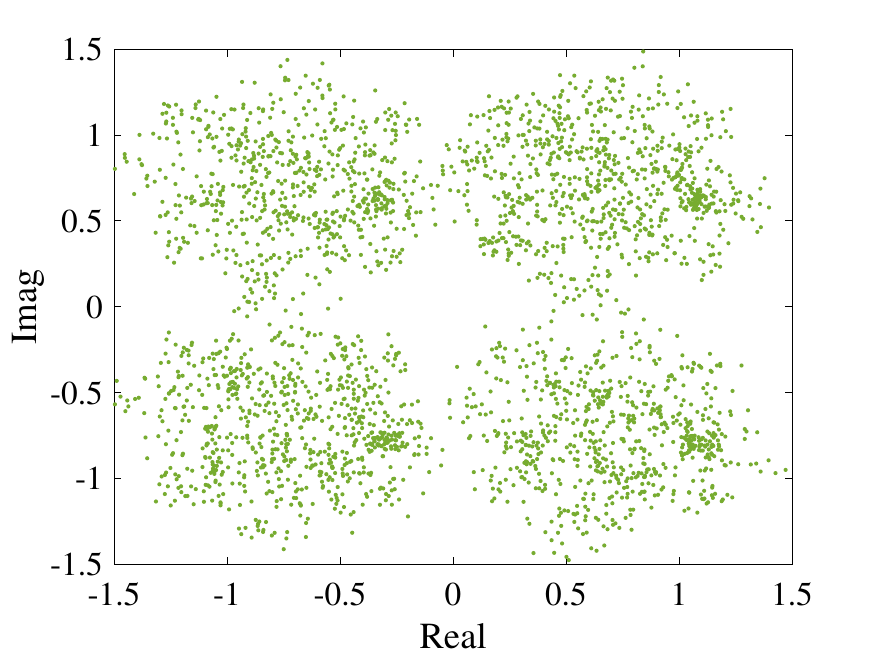}
	\label{fig:RISS_FF_align}}
	\hfill
	\subfloat[Robust interference suppression, $\Delta_\text{EVM}=20.0157\%$.]{\includegraphics[width=0.48\linewidth]{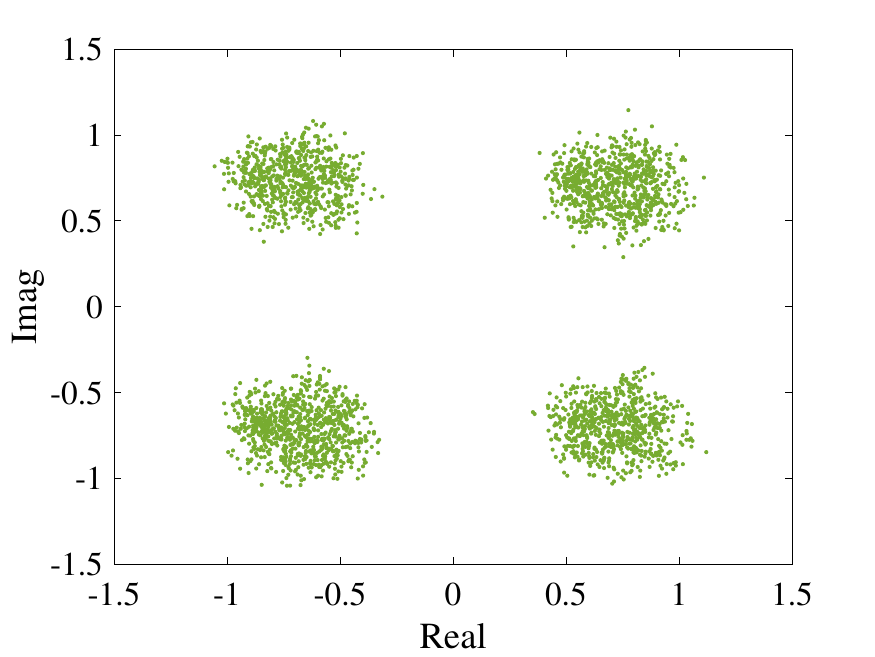}
	\label{fig:RISS_FF_interSupp}}
	\caption{Constellation diagram measurements of the Rx in far-field model.}
	\label{fig:constellation_FF_align_supp}
\end{figure}

\begin{figure}[!t]
	\centering
	\subfloat[Theoretical beam gain with 2-bit RISS.]{\includegraphics[width=0.85\linewidth]{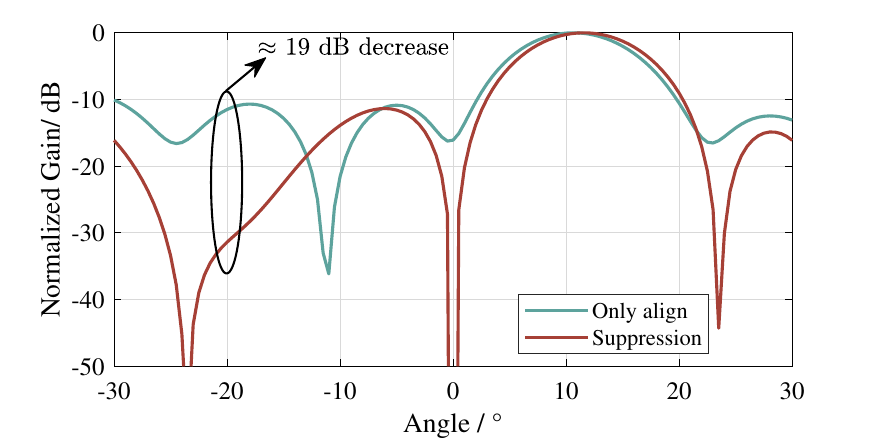}}
	\hfill
	\subfloat[Practical experimental beam gain.]{\includegraphics[width=0.85\linewidth]{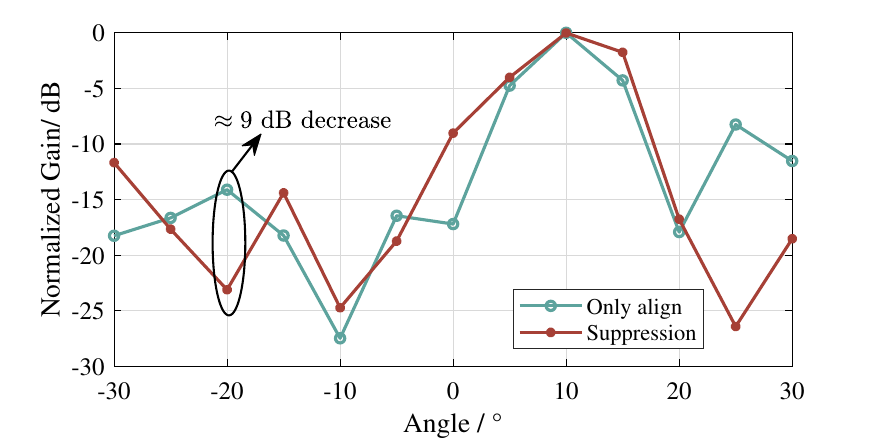}}
	\caption{The beampattern of align scheme and interference suppression scheme in the far-field model. The target signal is located in $\phi_2^\text{azi}=10^\circ$, while the interference signal is located in $\phi_1^\text{azi}=-20^\circ$.}
	\label{fig:beampattern_FF_align_supp}
\end{figure}

\begin{figure}
	\centering
	\subfloat[Theoretical heatmap of near-field scheme.]{\includegraphics[width=0.85\linewidth]{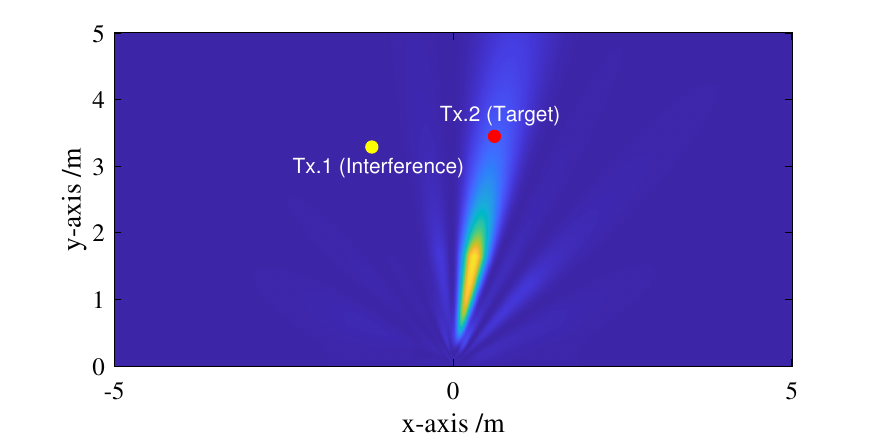}\label{fig:HP_NF_theo}}
	\hfill
	\subfloat[Theoretical heatmap of far-field scheme.]{\includegraphics[width=0.85\linewidth]{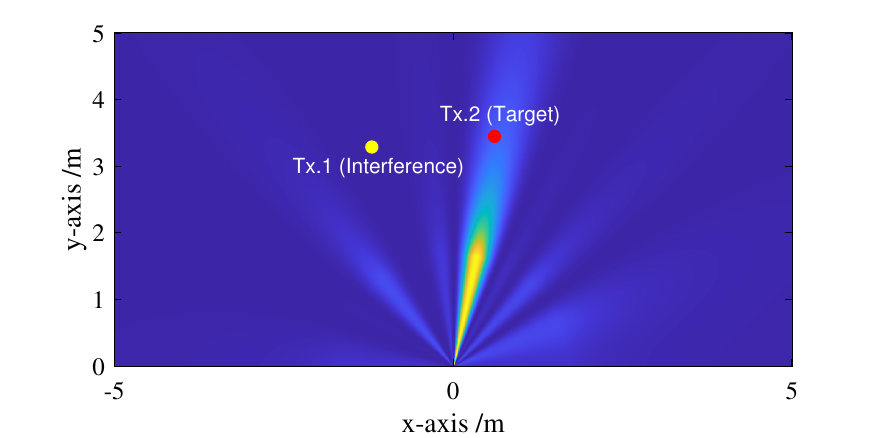}\label{fig:HP_FF_prac}}
	\caption{Heatmaps of both near-field and far-field schemes. The RISS is located at (0,0) and target signal is located in $\phi_2^\text{azi}=10^\circ$, while the interference signal is located in $\phi_1^\text{azi}=-20^\circ$.}
	\label{fig:2dheatmaps}
\end{figure}

In contrast, as illustrated in Fig. \ref{fig:constellation_FF_align_supp} and Fig. \ref{fig:beampattern_FF_align_supp}, the performance of target enhancement and interference suppression in the far-field model is weak. 
To better understand the underlying cause of this phenomenon, we present theoretical 2D heatmaps under both the near-field and far-field schemes. It can be observed from Fig. \ref{fig:2dheatmaps} that both schemes place the Tx.2 within the main-lobe while forming nulls in the directions of Tx.1. However, the near-field scheme exhibits fewer sidelobes and does not generate significant signal gain in the near region. In contrast, the far-field scheme produces more pronounced sidelobes and exhibits gain along the entire path at $\phi_2^\text{azi}=10^\circ$. In addition, there is a slight offset between the main-lobe direction and the actual position of the Tx.2. This misalignment is caused by the combined effects of the interference suppression constraint and the 2-bit RISS quantization. Furthermore, hardware imperfections, such as phase adjustment errors in the RISS, may also be contributing factors to the inferior performance of the far-field scheme compared to the near-field scheme.

\begin{figure}[!t]
	\centering
	\subfloat[Interference constellation with align scheme, $\Delta_\text{EVM}=24.5761\%$.]{\includegraphics[width=0.48\linewidth]{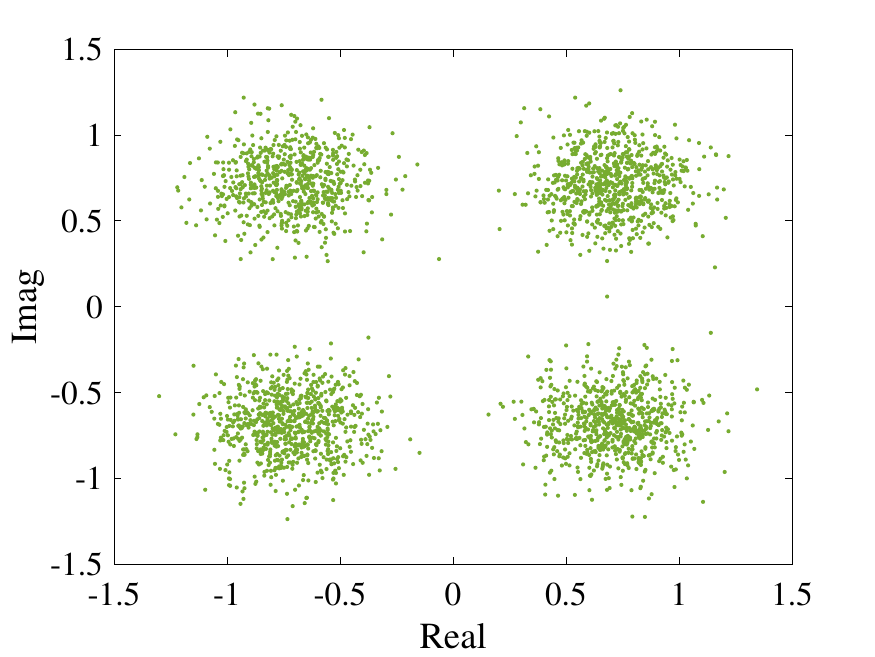}
	\label{fig:inter_align}}
	\hfill
	\subfloat[Interference constellation with robust interference suppression, $\Delta_\text{EVM}=59.7931\%$.]{\includegraphics[width=0.48\linewidth]{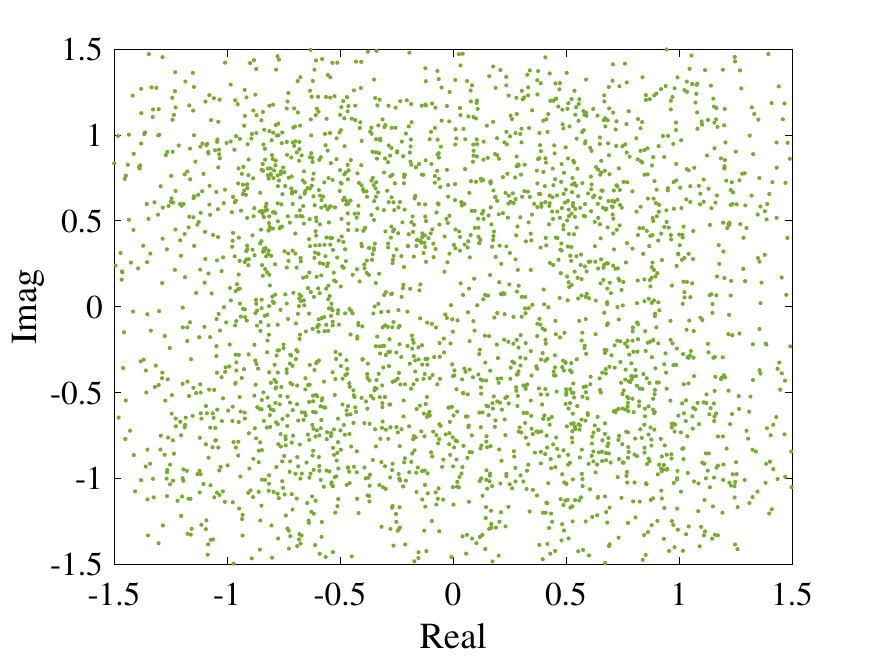}
	\label{fig:inter_supp}}
	\caption{Constellation diagram measurements of the Rx while only Tx. 2 is available.}
	\label{fig:iner_constellation_align_supp}
\end{figure}

Finally, to further investigate interference suppression, we deactivate Tx. 2, allowing the receiver to capture only the interference signal. The constellation measurements for both the align scheme and the interference suppression scheme using the near-field model are displayed in Fig. \ref{fig:iner_constellation_align_supp}. With only Tx. 1 active, the constellation in Fig. \ref{fig:inter_align} shows that the align scheme retains significant interference signal power. In contrast, the interference suppression scheme shows effective interference mitigation, resulting in a highly dispersed constellation diagram of the interference signal, further demonstrating the robustness of the interference suppression technique.

\section{Conclusion}\label{sec:VI}

This paper introduces a RISS-assisted communication system, with both far-field and near-field models designed for target signal enhancement and robust interference suppression. Leveraging the signal processing ability of active elements, the RISS operates independently, eliminating the need for traditional CSI feedback and enabling a purely forward transmission process. Extensive over-the-air experiments validate the system's sensing capabilities, such as DOA estimation and identification recognition, as well as the effectiveness of the RISS-based communication scheme for signal enhancement and robust interference suppression. While both far-field and near-field models are effective for target enhancement, the near-field model demonstrates superior performance in interference suppression within our experimental setup. The successful validation of this sensing-assisted communication system lays a solid foundation for future research on innovative communication schemes and advanced sensing beampattern designs.
{\appendices
\section{Convergence Analysis of the Proposed Algorithms}\label{app:A}
\subsection{Convergence analysis of the Proposed Algorithm 1}
By observing Eq. \eqref{eqn:eq15}, we note that $\boldsymbol{\mathcal{C}}_1$ and $\boldsymbol{\mathcal{C}}_2$ differ only by a constant term $P_i,\forall i\in\{1, 2\}$. For simplicity, we let $P_1 = P_2$, and denote the common term as $\boldsymbol{\mathcal{C}}$, i.e., $\boldsymbol{\mathcal{C}}_1^t = \boldsymbol{\mathcal{C}}_2^t = \boldsymbol{\mathcal{C}}_t$. For the proposed IRM algorithm, we begin by formulating the Lagrangian function corresponding to problem (P1.2) as follows
\begin{align} 
	\mathcal{L} &= -\text{trace}(\boldsymbol{\mathcal{C}}_t \boldsymbol{\mathcal{A}}_d) + \epsilon_t r + \boldsymbol{\mu}^T (\text{diag}(\boldsymbol{\mathcal{C}}_t) - M\mathbf{1}_N) \nonumber\\ 
	&\qquad\qquad + \sum_{k \neq d} \lambda_k \left(\text{trace}(\boldsymbol{\mathcal{C}}_t \boldsymbol{\mathcal{A}}_k) - \tau_k\right)\nonumber\\
	&\qquad\qquad  - \text{trace}\left(\boldsymbol{\Gamma}_1\left(r \mathbf{I}_{N-1}- \mathbf{V}_t^H \boldsymbol{\mathcal{C}}_t \mathbf{V}_t \right)\right)  \nonumber\\ 
	&\qquad\qquad - \text{trace}\left(\boldsymbol{\Gamma}_2 \boldsymbol{\mathcal{C}}_t \right) + \lambda_e \left(\text{trace}(\boldsymbol{\mathcal{C}}_t) - MN^2\right), 
\end{align}
where $\boldsymbol{\mu} \in \mathbb{R}^N$, $\lambda_k \geq 0$, $\boldsymbol{\Gamma}_1 \in \mathbb{S}_+^{N-1}$, $\boldsymbol{\Gamma}_2 \in \mathbb{S}_+^{N}$, and $\lambda_e \geq 0$ denote the dual multipliers associated with the equality constraints, inequality constraints, semidefinite constraints, and trace constraint, respectively. $\mathbb{R}^N$ denotes the space of $N$-dimensional real-valued vectors, and $\mathbb{S}_+^N$ denotes the cone of $N \times N$ real symmetric positive semidefinite matrices.

Rewriting the Lagrangian in a more compact form yields
\begin{align}
	\mathcal{L} = &\text{trace}\left( \boldsymbol{\mathcal{C}}_t \boldsymbol{\Pi} \right) + r (\epsilon_t - \text{trace}(\boldsymbol{\Gamma}_1)) - \boldsymbol{\mu}^T (M\mathbf{1}_N)\nonumber\\
	&\qquad\qquad\qquad\qquad  - \sum_{k \neq d} \lambda_k \tau_k - \lambda_e MN^2,
\end{align}
where $\boldsymbol{\Pi} = -\boldsymbol{\mathcal{A}}_d + \text{diag}(\boldsymbol{\mu}) + \sum_{k \neq d} \lambda_k \boldsymbol{\mathcal{A}}_k + \mathbf{V}_t \boldsymbol{\Gamma}_1^H \mathbf{V}_t^H - \boldsymbol{\Gamma}_2 + \lambda_e \mathbf{I}_N$. The dual function is then defined as $g = \inf_{\boldsymbol{\mathcal{C}}_t, r} \mathcal{L}$, which is finite if the stationarity conditions are satisfied. The first-order optimality conditions are given by
\begin{align} 
	&\frac{\partial \mathcal{L}}{\partial r} = \epsilon_t - \text{trace}(\boldsymbol{\Gamma}_1) = 0, \nonumber\\ 
	&\frac{\partial \mathcal{L}}{\partial \boldsymbol{\mathcal{C}}_t} = \boldsymbol{\Pi} = \mathbf{0}. 
\end{align}

Thus, the dual problem can thus be written as
\begin{align} 
	\min_{\boldsymbol{\mu}, \boldsymbol{\lambda}, \boldsymbol{\Gamma}_1, \boldsymbol{\Gamma}_2, \lambda_e} \quad & \boldsymbol{\mu}^T (M\mathbf{1}_N) + \sum_{k \neq d} \lambda_k \tau_k + \lambda_e MN^2 \label{eqn:appA_dualPro}\\ 
	\text{s.t.} \quad & \boldsymbol{\Pi} = \mathbf{0}, \tag{\ref{eqn:appA_dualPro}a}\\ 
	& \text{trace}(\boldsymbol{\Gamma}_1) = \epsilon_t, \tag{\ref{eqn:appA_dualPro}b}\\ 
	& \lambda_k \geq 0, \boldsymbol{\Gamma}_1 \succeq \mathbf{0}, \boldsymbol{\Gamma}_2 \succeq \mathbf{0}, \lambda_e \geq 0\tag{\ref{eqn:appA_dualPro}c}. 
\end{align}

Since $\boldsymbol{\mu}$ corresponds to equality constraints, it is not sign-restricted. By inspecting the primal problem, it can be verified that it is convex and satisfies Slater’s condition. Therefore, strong duality holds.

Let $(\boldsymbol{\mathcal{C}}_t^*, r_t^*)$ and $(\boldsymbol{\mathcal{C}}_{t+1}^*, r_{t+1}^*)$ denote the optimal solutions at iterations $t$ and $t+1$, respectively. The difference in the primal objective function values is given by
\begin{align} 
	p_{t+1}^* - p_t^* &= \left( -\text{trace}(\boldsymbol{\mathcal{C}}_{t+1}^* \boldsymbol{\mathcal{A}}_d) + \epsilon_{t+1} r_{t+1}^* \right)\nonumber\\ 
	&\qquad\qquad\qquad\qquad - \left(-\text{trace}(\boldsymbol{\mathcal{C}}_t^* \boldsymbol{\mathcal{A}}_d) + \epsilon_t r_t^* \right) \nonumber\\ 
	&= -\text{trace}\left( (\boldsymbol{\mathcal{C}}_{t+1}^* - \boldsymbol{\mathcal{C}}_t^*) \boldsymbol{\mathcal{A}}_d \right) + \epsilon_{t+1} r_{t+1}^* - \epsilon_t r_t^*. 
\end{align}

By substituting $\epsilon_{t+1} = \epsilon_t \varsigma$ and applying strong duality, i.e., $p_{t+1}^* - p_t^* = d_{t+1}^* - d_t^*$, we obtain
\begin{align}
	\epsilon_t \varsigma r_{t+1}^* - \epsilon_t r_t^* = \underbrace{(d_{t+1}^* - d_t^*)}_{\Delta_d} + \underbrace{\text{trace}\left( (\boldsymbol{\mathcal{C}}_{t+1}^* - \boldsymbol{\mathcal{C}}_t^*) \boldsymbol{\mathcal{A}}_d \right)}_{\Delta_c},
\end{align}
where both $\Delta_d$ and $\Delta_c$ are bounded due to the compactness of the feasible set. Rearranging the expression leads to
\begin{align}
	\varsigma r_{t+1}^* - r_t^* = \frac{\Delta_d + \Delta_c}{\epsilon_t}.
\end{align}

Taking the limit as $t \to \infty$, and using the fact that $\epsilon_t = \epsilon_0 \varsigma^t \to \infty$, we obtain
\begin{align}
	\lim_{t \to \infty} \left( \varsigma r_{t+1}^* - r_t^* \right) = 0.
\end{align}

Since $r_t^* \to 0$ due to feasibility, the asymptotic behavior of the sequence satisfies
\begin{align}
	r_{t+1}^* \approx \frac{r_t^*}{\varsigma}.
\end{align}

Therefore, $r_t^*$ converges to zero at a linear rate of $\frac{1}{\varsigma}$.

Having established that $r_t^*$ converges to zero at a linear rate of $1/\varsigma$, we next show that the sequence ${\boldsymbol{\mathcal{C}}_t^*}$ converges to a locally optimal solution $\boldsymbol{\mathcal{C}}^*$, which satisfies $\text{rank}(\boldsymbol{\mathcal{C}}^*) \leq 1$. Note that this conclusion aligns with the results in \cite[Proposition 3.6]{IRMalg}.

Specifically, by the constraint $(c)$ and the fact that $\lim_{t \to \infty} r_t^* = 0$, we have
\begin{align}
	r_t^* \mathbf{I}_{N-1} - \mathbf{V}_t^H \boldsymbol{\mathcal{C}}_t^* \mathbf{V}_t \succeq \mathbf{0} \quad \Rightarrow \quad \lim_{t \to \infty} \mathbf{V}_t^H \boldsymbol{\mathcal{C}}_t^* \mathbf{V}_t = \mathbf{0}.
\end{align}

Since $\mathbf{V}_t \in \mathbb{C}^{N \times (N-1)}$ is full column-rank, it follows that$\text{rank}(\boldsymbol{\mathcal{C}}_t^*) \leq 1.$ Thus, the optimal solution at iteration $t$ can be written in the form
\begin{align}
	\boldsymbol{\mathcal{C}}_t^* = \alpha_t \mathbf{c}_t \mathbf{c}_t^H, \quad \alpha_t \geq 0, \quad \|\mathbf{c}_t\|_2 = 1,
\end{align}
where $\mathbf{c}_t$ is the normalized dominant eigenvector of $\boldsymbol{\mathcal{C}}_t^*$.

Now, consider the difference in the constraint projections between successive iterations
\begin{align}
	\lim_{t \to \infty} \left( \mathbf{V}_{t+1}^H \boldsymbol{\mathcal{C}}_{t+1}^* \mathbf{V}_{t+1} - \mathbf{V}_t^H \boldsymbol{\mathcal{C}}_t^* \mathbf{V}_t \right) = \mathbf{0}.
\end{align}

Since $\text{rank}(\boldsymbol{\mathcal{C}}_t^*) = 1$ asymptotically, we may assume
\begin{align}
	\boldsymbol{\mathcal{C}}_{t+1}^* \approx \alpha \boldsymbol{\mathcal{C}}_t^*, \quad \alpha \in \mathbb{R},
\end{align}
which implies that $\boldsymbol{\mathcal{C}}_{t+1}^*$ and $\boldsymbol{\mathcal{C}}_t^*$ share the same eigenvectors. Hence, $\lim_{t \to \infty} \mathbf{c}_{t+1} = \mathbf{c}_t.$ As a consequence, the basis $\mathbf{V}_t$ becomes asymptotically constant, i.e., $\mathbf{V}_t \to \mathbf{V}^*$ as $t \to \infty$. Therefore, under the fixed $\mathbf{V}^*$, the subproblem (P1.2) converges to a locally optimal solution $\boldsymbol{\mathcal{C}}^*$ that satisfies the Karush-Kuhn-Tucker conditions (KKT) conditions.

\subsection{Convergence Analysis of the Proposed Algorithm 2}
Note that the proposed AO algorithm can be regarded as a special case of the block coordinate descent (BCD) method. Specifically, the phase vector of the RISS, denoted as $\boldsymbol{\varphi} = \{\varphi_1, \dots, \varphi_N\},\varphi_k \in [0, 2\pi),\forall k \in N$, is naturally partitioned into $N$ scalar blocks. In each iteration of the AO algorithm, only one scalar variable $\varphi_k$ is optimized while keeping all other variables fixed, which aligns precisely with the coordinate-wise update rule in the BCD framework. Therefore, the proposed AO algorithm can be interpreted as a BCD method with scalar block updates and exact maximization along each coordinate. This observation enables us to leverage classical results from BCD theory to rigorously yet concisely support the convergence analysis.

The convergence of the AO algorithm is ensured by the following three properties
\begin{itemize}
	\item Compactness and boundedness: The domain of $\boldsymbol{\varphi}$ is compact, i.e., $\varphi_k\in[0, 2\pi), \forall k\in N$, and the objective function $f(\boldsymbol{\varphi})=S_2-\eta S_1$ is continuous and upper-bounded over this domain. Specifically, $f(\boldsymbol{\varphi}) \leq N$, ensuring the existence of a finite supremum.
	
	\item Monotonic ascent: In each iteration, the objective function is exactly maximized over a single coordinate using a one-dimensional search. This guarantees that the objective value is non-decreasing across iterations, i.e., $f(\boldsymbol{\varphi}^{(t+1)}) \geq f(\boldsymbol{\varphi}^{(t)})$.
	
	\item Convergence to a stationary point: Since the sequence ${f(\boldsymbol{\varphi}^{(t)})}$ is monotonically non-decreasing and upper-bounded, it converges to a finite limit $f^*$. Moreover, any limit point $\boldsymbol{\varphi}^* = \{\varphi_1^*, \dots, \varphi_N^*\}$ satisfies coordinate-wise optimality. That is, further improvement in the objective function is not possible by updating any individual coordinate $\varphi_k$. When the objective function is differentiable, this implies that the gradient vanishes along each coordinate direction, indicating that $\boldsymbol{\varphi}^*$ is a stationary point.
\end{itemize}
Based on these three conditions and standard results from BCD convergence theory\cite{tseng2001convergence}, we conclude that the proposed algorithm converges to a stationary point. For a more detailed derivation, readers are referred to classical convergence proofs of BCD algorithms\cite{tseng2001convergence}. 
\section{Hardware Information Supplement}\label{app:B}
The detailed parameters of RIS can be found in Table \ref{table:para}. 

\begin{table}[]
	\centering
	\caption{Product specifications for the 3.5 GHz RIS from Beijing Actenna Technology Co., Ltd.}
	\begin{tabular}
		{|>{\bfseries}c|>{\itshape}c|}
		\hline
		\multicolumn{2}{|c|}{\textbf{Product Specifications}} \\
		\hline
		Product Model & I24-S35 \\ 
		\hline
		Working Frequency & 3.4--3.6 GHz \\
		\hline
		Array Size & 10$\times$10 units \\
		\hline
		Phase-Shifting Step & 2-bit \\
		\hline
		Phase Accuracy & 90$^{\circ}$$\pm$15$^{\circ}$ \\
		\hline
		Insertion Loss & 1.0--3.0 dB \\
		\hline
		Polarization Mode & Orthogonal dual-linear polarization \\
		\hline
		Angle Range & $\pm$60$^{\circ}$ (H/V) \\
		\hline
		Power Consumption & 9W (Typical) \\
		\hline
		Control Port & DB9 \\
		\hline
		Input Voltage & 9--36V DC \\
		\hline
		Dimensions & 430 mm (W) $\times$ 430 mm (H) $\times$ 46 mm (D) \\
		\hline
		Weight & 3.0 kg \\
		\hline
	\end{tabular}\label{table:para}
\end{table}

Moreover, Fig. \ref{fig:ant} illustrate the three-dimensional structural model of the sensing antenna array, which consists of four rectangular microstrip patch antennas arranged in parallel, with a back-fed configuration. To mitigate antenna coupling and address other design considerations, the spacing between the units is set to 0.0590 meters, corresponding to 0.6883 wavelengths at 3.5 GHz. The physical dimensions of the entire antenna array are 236 mm × 59 mm × 9.2 mm.

\begin{figure}[!t]
	\centering
	{
		\subfloat[]{\includegraphics[width = 0.5\linewidth]{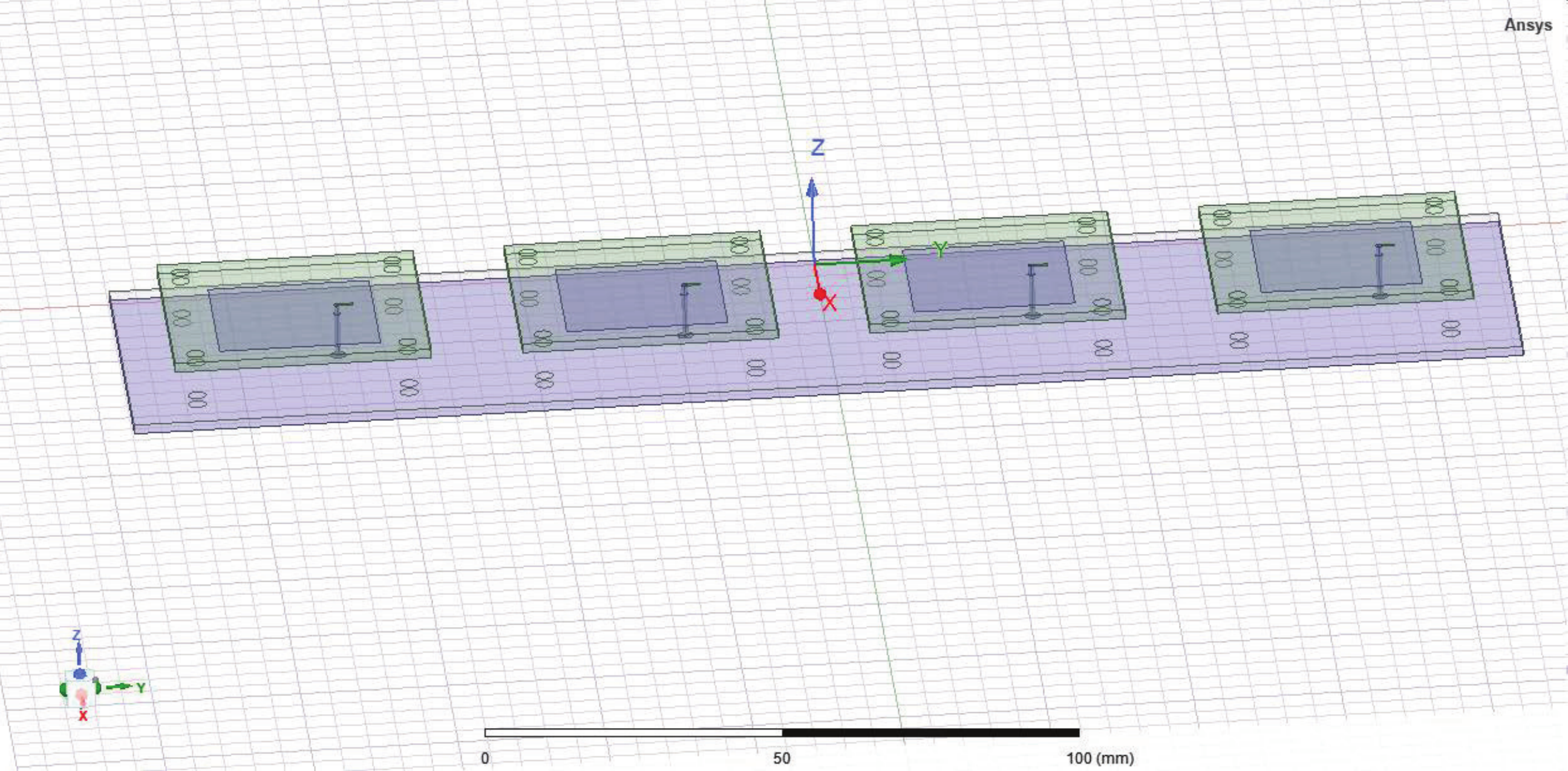}}
		\hfil
		\subfloat[]{\includegraphics[width = 0.48\linewidth]{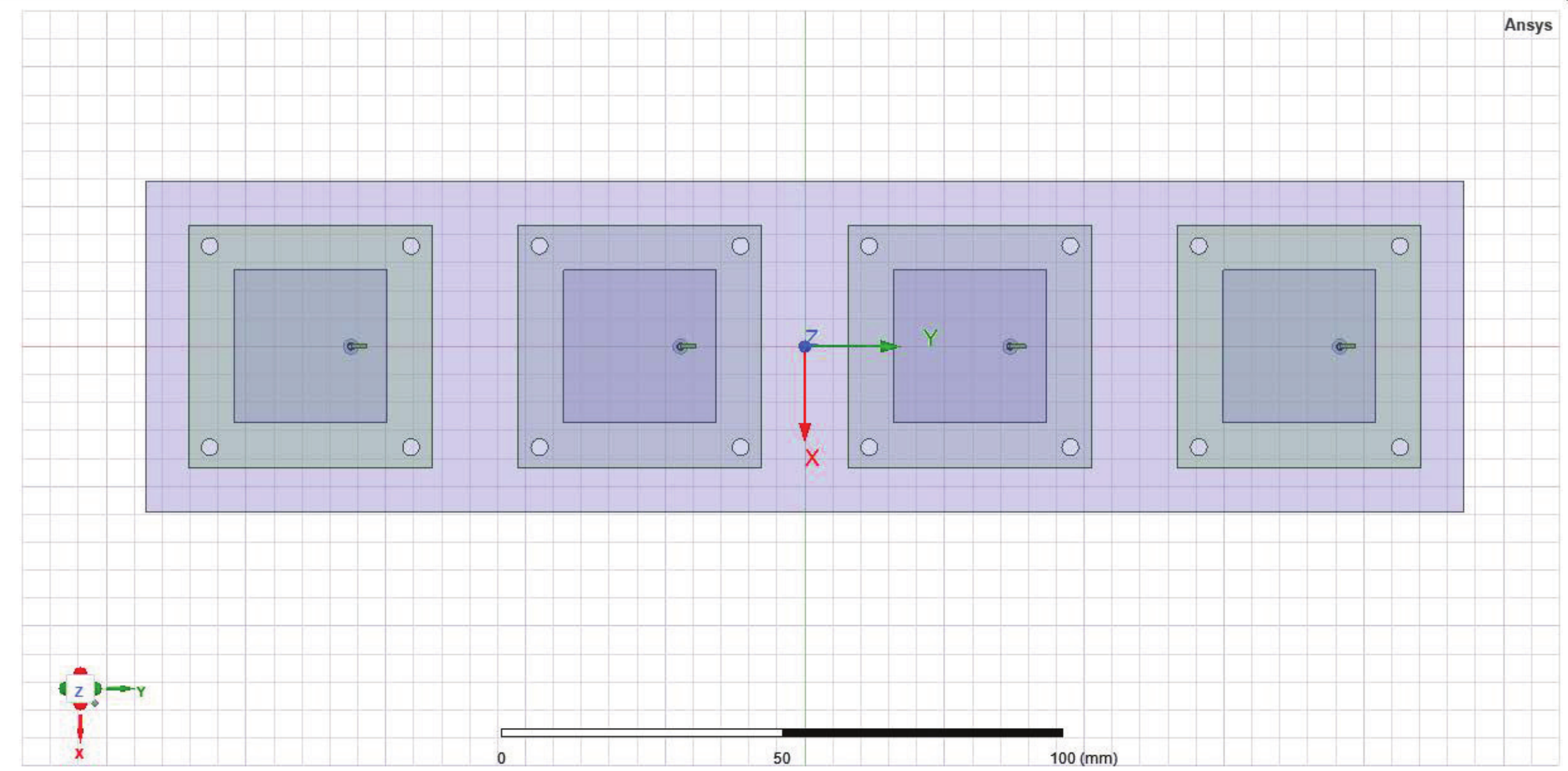}}
	}
	\caption{Three-dimensional structural model of the sensing antenna array.}
	\label{fig:ant}
\end{figure}

\begin{figure}
    \centering
	\includegraphics[width=0.80\linewidth]{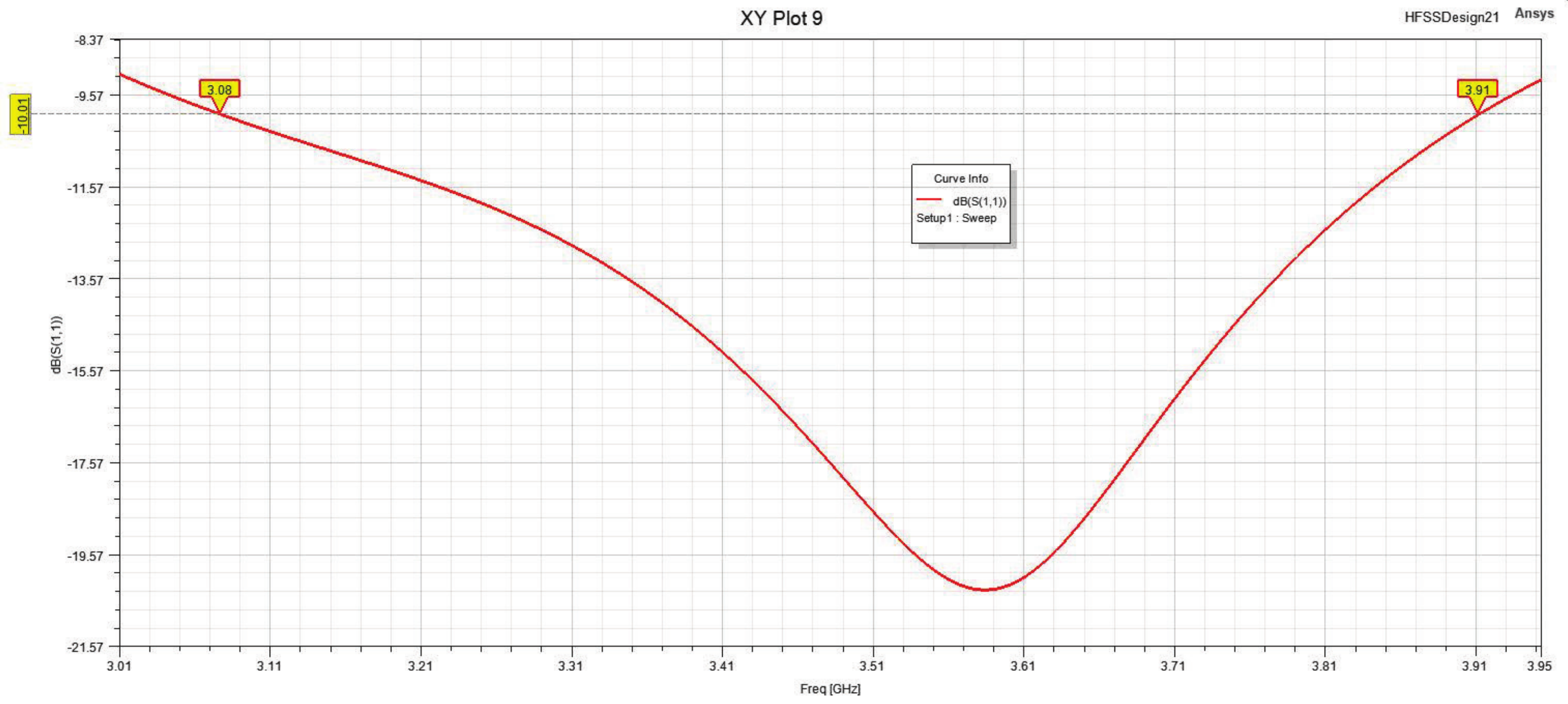}
	\caption{$S_{11}$ reflection coefficient curve of sensing antenna array.}
	\label{fig:S11}
\end{figure}

Fig. \ref{fig:S11} shows the $S_{11}$ reflection coefficient curve. Using a reference of -10 dB, which corresponds to a reflection loss of less than or equal to 10\%, the antenna array exhibits a bandwidth ranging from approximately 3.09 GHz to 3.91 GHz, resulting in a total bandwidth of about 830 MHz. The resonance point is observed around 3.58 GHz, suggesting that the antenna array is capable of operating effectively at 3.5 GHz, as discussed in this paper.

\begin{figure}
    \centering
	\includegraphics[width=0.99\linewidth]{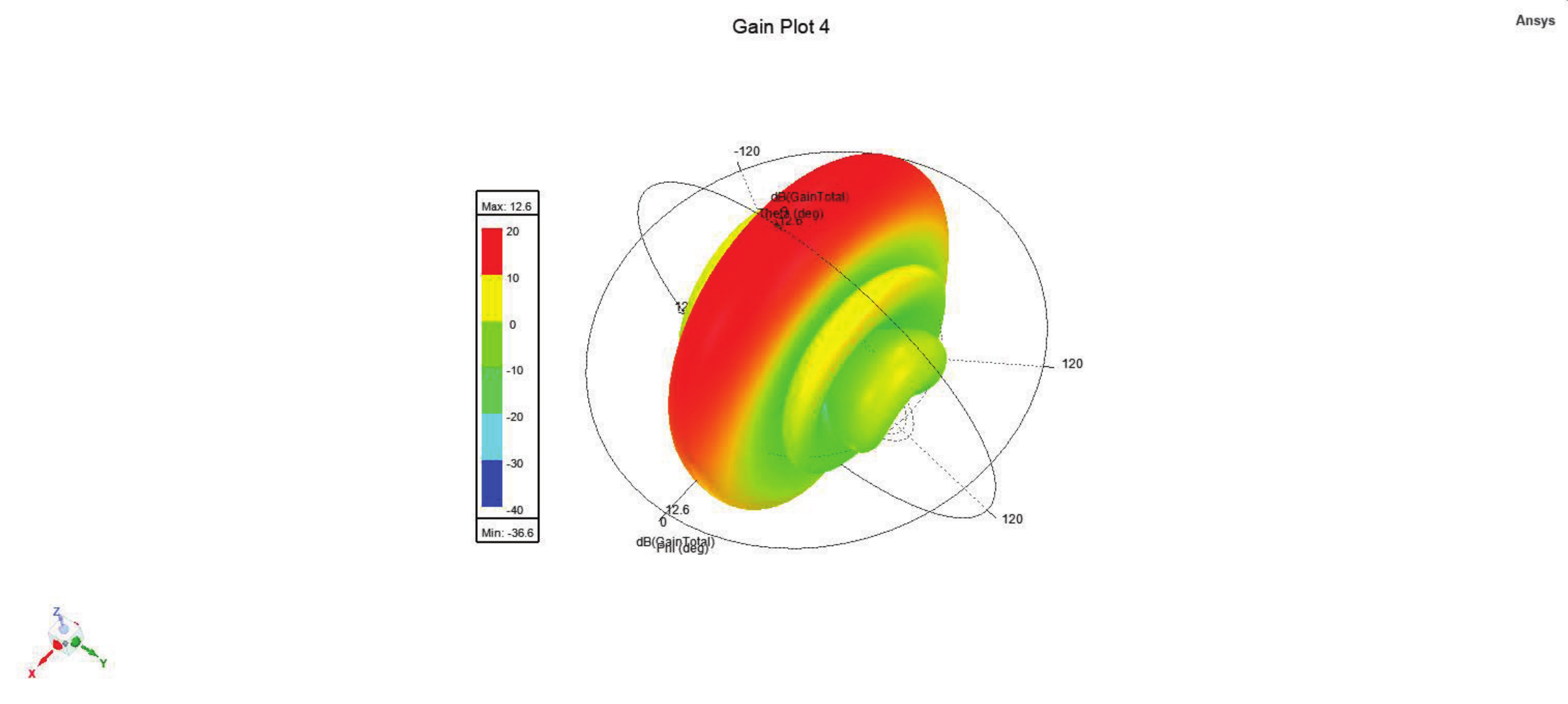}
	\caption{The beampattern of sensing antenna array.}
	\label{fig:beampattern}
\end{figure}

Fig. \ref{fig:beampattern} presents the radiation pattern of the sensing antenna array at an operating frequency of 3.5 GHz. The measured maximum gain reaches 12.6 dB, demonstrating strong directional radiation capability within the main lobe. As illustrated in Fig. \ref{fig:ant}, the $y$-axis is defined as the horizontal axis in this configuration, with horizontal linear polarization aligned along the $y$-axis direction.
}
\bibliography{Reference}
\end{document}